\documentclass[useAMS,usenatbib,10pt]{mn2e}

\usepackage{graphicx, natbib,epsfig,amsmath,amssymb, mathrsfs, txfonts} %
\DeclareGraphicsExtensions{.eps, .jpg, .ps}

\usepackage{epstopdf}

\title[Laboratory comparison of coronagraphic concepts under dynamical seeing and XAO correction]
  {Laboratory comparison of coronagraphic concepts under dynamical seeing and high-order adaptive optics correction}
\author[P. Martinez et al.]
  {P.~Martinez,$^{1, 2}$ \thanks{Now at IPAG: patrice.martinez@obs.ujf-grenoble.fr}
  E.~Aller-Carpentier,$^1$ M.~Kasper,$^1$ A.~Boccaletti,$^3$ C.~Dorrer,$^4$ J.~Baudrand$^3$
 \\
  $^1$ European Southern Observatory, Karl-Schwarzschild-Strasse 2, D-85748, Garching, Germany \\
  $^2$ UJF-Grenoble 1 / CNRS-INSU, Institut de Plan\'{e}tologie et d'Astrophysique de Grenoble (IPAG) UMR 5274, Grenoble, F-38041, France \\  
  $^3$ LESIA, Observatoire de Paris Meudon, 5 pl. J. Janssen, 92195 Meudon, France \\
  $^4$ Aktiwave, 241 Ashley drive, Rochester, NY, 14620, USA}
\date{Released 2011 February 11}

\pagerange{\pageref{firstpage}--\pageref{lastpage}} \pubyear{2011}

\begin{document}

\label{firstpage}

\maketitle

\begin{abstract}
The exoplanetary science through direct imaging and spectroscopy will largely expand with the forthcoming development of new instruments at the VLT (SPHERE), Gemini (GPI), Subaru (HiCIAO), and Palomar (Project 1640) observatories. 
All these ground-based adaptive optics instruments combine extremely high performance adaptive optics (XAO) systems correcting for the atmospheric turbulence with advanced starlight-cancellation techniques such as coronagraphy to deliver contrast ratios of about $10^{-6}$ to $10^{-7}$.
While the past fifteen years have seen intensive research and the development of high-contrast coronagraph concepts, very few concepts have been tested under dynamical seeing conditions (either during sky observation or in a realistic laboratory environment). 
In this paper, we discuss the results obtained with four different coronagraphs -- phase and amplitude types -- on the High-Order Testbench (HOT), the adaptive optics facility developed at ESO. This facility emphasizes realistic conditions encountered at a telescope (e.g., VLT), including a turbulence generator and a high-order adaptive optics system. It enables to evaluate the performance of high-contrast coronagraphs in the near-IR operating with an AO-corrected PSF of 90$\%$ Strehl ratio under 0.5$\arcsec$ dynamical seeing.

\end{abstract}


\section{Introduction}
The imagery and spectroscopy of extrasolar planets are among the most exciting and ambitious goals of contemporary observational programs. 
Direct detection and characterization of faint objects around bright astrophysical sources is highly challenging due to the large flux ratio and small angular separation. 
For instance, self-luminous giant planets are typically $10^{6}$ times fainter than their parent star in the near-infrared.
In this context, the worldwide emergence of high-contrast imaging instruments, e.g., SPHERE \citep{SPHERE}, GPI \citep{GPI}, HiCIAO \citep{HICIAO}, or Project 1640 \citep{Oppenheimer04, hinkley11} will have the potential to dramatically enlarge the actual discovery space of exo-planets. 
These instruments will use extreme adaptive optics (XAO) systems in association with coronagraphy to overcome the contrast issue.

A coronagraph used in association with an AO system can improve the sensitivity of an imaging system to faint structures surrounding a bright source. 
These devices block the core of the image of an on-axis source and suppress the bright diffraction rings and halo that would otherwise reduce the dynamic range. 
The state-of-the-art of coronagraphy has impressively expanded during the past fifteen years,  motivated by the detection and imaging of exoplanets, ideally down to Earth-like planets. 
An extensive review of the different families of coronagraph was carried out by \citet{guyon06}, where optimal solutions were proposed in the context of space-based observations. Likewise, for ground-based instruments several concepts have been compared \citep{martinez08}.

Coronagraphs can now provide a very large on-axis extinction as demonstrated in laboratory conditions \citep[e.g.,][]{riaud03, Abe03, mawet06,  creep06,  enya07, enya08, mawet09,  martinez09a, guyon10, mamadou10, martinez10a}, while very few have been tested on-sky \citep[e.g.,][]{baudoz00, Boccaletti04, swartzlander08, mawet10}. Their performance are impacted by the large amount of residual phase aberrations that are left uncorrected by the AO system.  Although coronagraphy is a mandatory technique, this critical subsystem can only reduce the contribution of the coherent part of the on-axis starlight. As a result, coronagraphic on-sky capabilities strongly depend on the AO system performance.

In this context and in the framework of the SPHERE and EPICS \citep[][for the future European-extremely large telescope]{EPICS} instruments, we have developed and tested four different coronagraphic concepts: a four-quadrants phase mask \citep{rouan00}, several Lyot coronagraphs \citep{Lyot39}, an apodized-pupil Lyot coronagraph \citep{aime02, soummer03}, and a band-limited coronagraph \citep{kuchner02}. The objective is to compare their respective behavior under realistic conditions and to analyze their responses to the XAO system. The selection of an optimal concept is a byproduct of this study as initiated in a previous study by means of simulations \citep{martinez08}. These four prototypes were tested on HOT, the High-Order Testbench  \citep{vernet06, carpentier08}, which 
reproduces realistic conditions at a telescope (e.g., VLT), including a turbulence generator, and a high-order adaptive optics system. HOT provides a practical realistic environment (e.g., Strehl ratio of 90$\%$ in H-band under 0.5$\arcsec$ dynamical seeing) to assess coronagraphic performance in the realm of high-contrast imaging instruments. Therefore, this study intends to provide comparative insights of coronagraphic devices but cannot address performance estimation of a particular concept for a real instrument. Although HOT provides a realistic environment  for evaluating high-contrast techniques, it cannot deliver similar design optimization and stability, nor operational conditions (i.e., observational and data reduction strategies, instrumental speckle suppression techniques, etc...) provided by real instruments currently in use (Project 1640) or being commissioned (SPHERE, GPI, or HiCIAO).

In Sect. 2 we present the optical setup and experimental conditions, while in Sect. 3 all coronagraphic concepts are described and details of their performance characterization are provided. In Sect. 4 we discuss the laboratory results obtained under dynamical seeing and XAO, and finally in Sect. 5 we draw conclusions.

\section{Optical setups}
In the following we present two optical setups. The first one, the \textit{High-Order Testbench} is used to test coronagraphs under realistic conditions with dynamical seeing and XAO correction, while the second one, the \textit{Coronagraphic Testbench}, is a dedicated optical setup for the characterization of the coronagraph prototypes before their implementation and use on HOT (i.e., to assess their intrinsic limitations). The latter corresponds to the near-IR arm of HOT assuming minor modifications. 

\subsection{High-Order Testbench}
\begin{figure*}
\includegraphics[width=12.8cm]{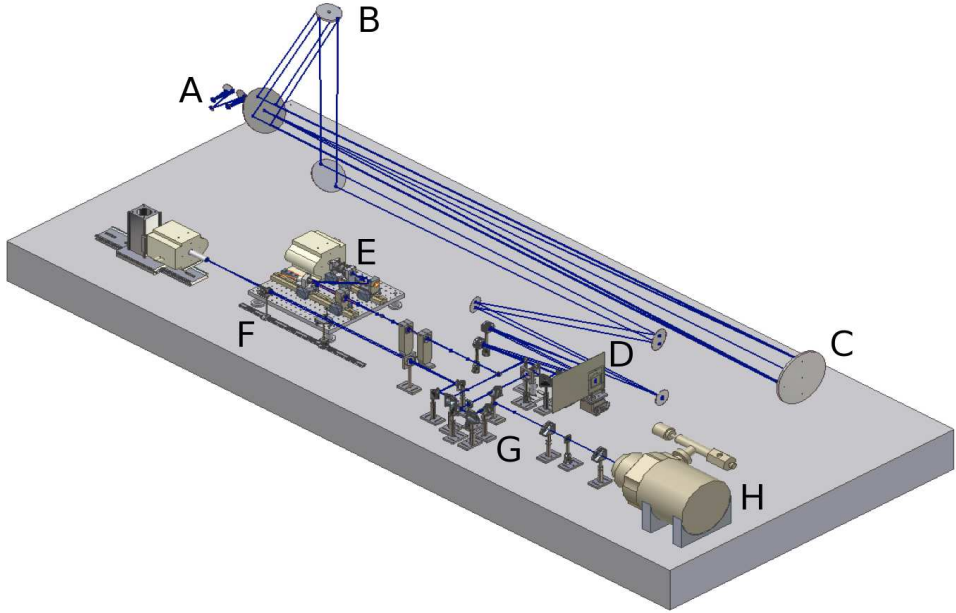}
\caption{The High-Order Testbench optical setup: turbulence generator (A), VLT-pupil, tip-tilt mount and bimorph-DM (B), spherical on-axis mirror (C), deformable mirror (D), Pyramid wavefront sensor (E), Shack-Hartmann wavefront sensor (F), science channel (near-IR path, G), and infrared camera (H).}
\label{BENCH}
\end{figure*}

The High-Order Testbench (HOT) is a high-contrast imaging adaptive optics bench installed at the ESO headquarters. It implements an XAO system and a star and turbulence generator to create realistic conditions encountered at a telescope. 
It provides ideal conditions to study XAO and coronagraphy.    
HOT gathers several critical components as shown in Fig. \ref{BENCH}: a turbulence generator with phase screens to simulate real seeing conditions (A), a VLT-pupil mask installed on a Tip-Tilt mount, and a 60-bimorph large stroke deformable mirror (B), a 32$\times$32 micro deformable mirror (D) -- DM, electrostatic MEMS device --, a beam splitter transmitting the visible light to wavefront sensing either with a pyramid concept (PWFS, E), or a Shack-Hartmann (SHWFS, F), while the infrared light is directed towards a coronagraph (G) and an infrared camera (1k$\times$1k HAWAII detector, H), and the ESO SPARTA real-time computer (RTC). All the optics are set on a table with air suspension in a dark room and are fully covered with protection panels forming a nearly closed box. 
After the generation of the dynamical aberrations, the output f/16.8 beam is transformed into an f/51.8 beam by a spherical on-axis mirror (C), and directed towards the pupil plane located about 1010 mm above the table level, with its axis tilted at 13.26 degrees as in the VLT Coude train. After that, relay optics prior to the beam splitter use flat mirrors and spherical mirrors to produce an f/50 telecentric beam. All relayed optics in the IR-path are made with IR achromatic doublets.  

The Shack-Hartmann wavefront sensor (F), developed by the University of Durham, provides a plate scale of 0.5$\arcsec$/pixel with 31x31subapertures, each ones sampled by 4x4 pixels of a 24 $\mu$m pixel size L3-CCD (Andor camera, read-out noise $<$ 1e-). The SHWFS real-time computer is an all-CPU architecture.

While not used during the experiment, HOT contains a Pyramid wavefront sensor built by Arcetri \citep{pinna08}, which is formally equivalent to the LBT wavefront sensor optical design, and consists in a double refractive pyramid modulated by a tip-tilt mirror, combined with an L3CCD-camera. The PWFS uses a dedicated all-CPU RTC.

The input source for HOT is a white-light source combined either with an IR narrow-band filter of $\Delta \lambda/\lambda$ = 1.4$\%$, central wavelength of 1.64 $\mu$m, or a broadband H filter centered on 1.6 $\mu$m, $\Delta \lambda/\lambda$ = 24$\%$, fed in by an 8 $\mu$m fiber. 

In the science arm of HOT (IR-path), a pupil-imager system is implemented for coronagraphic components alignment purpose.
All pupil-plane coronagraphic components are placed into opto-mechanical mounts that allow x-, y-, focus and rotation adjustment, while focal-plane masks can be adjusted in x-, y-, focus and tip-tilt.
 
\subsection{Coronagraphic Testbench}
\begin{figure}
\centering
\includegraphics[width=8.0cm]{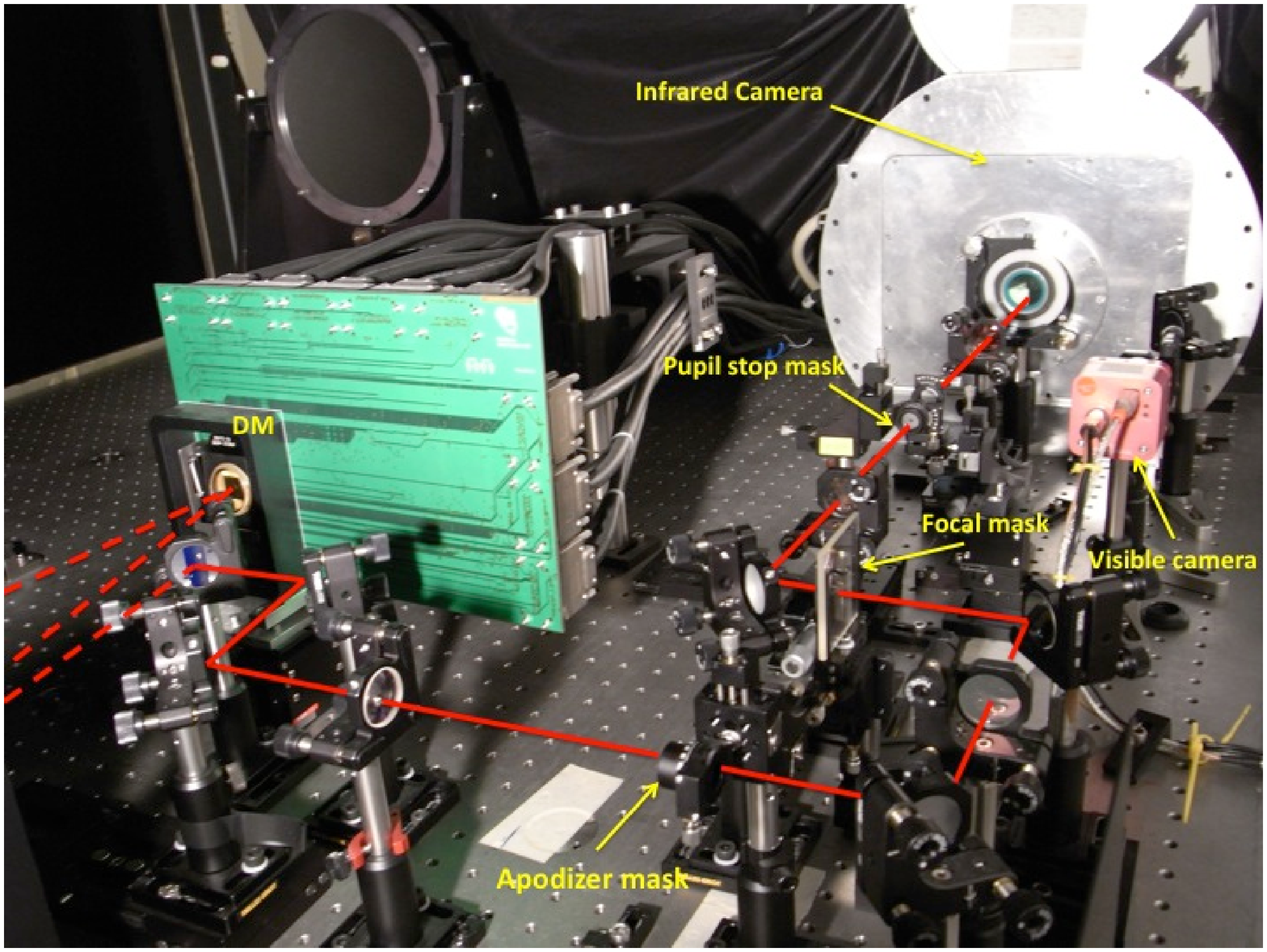}
\includegraphics[width=8.0cm]{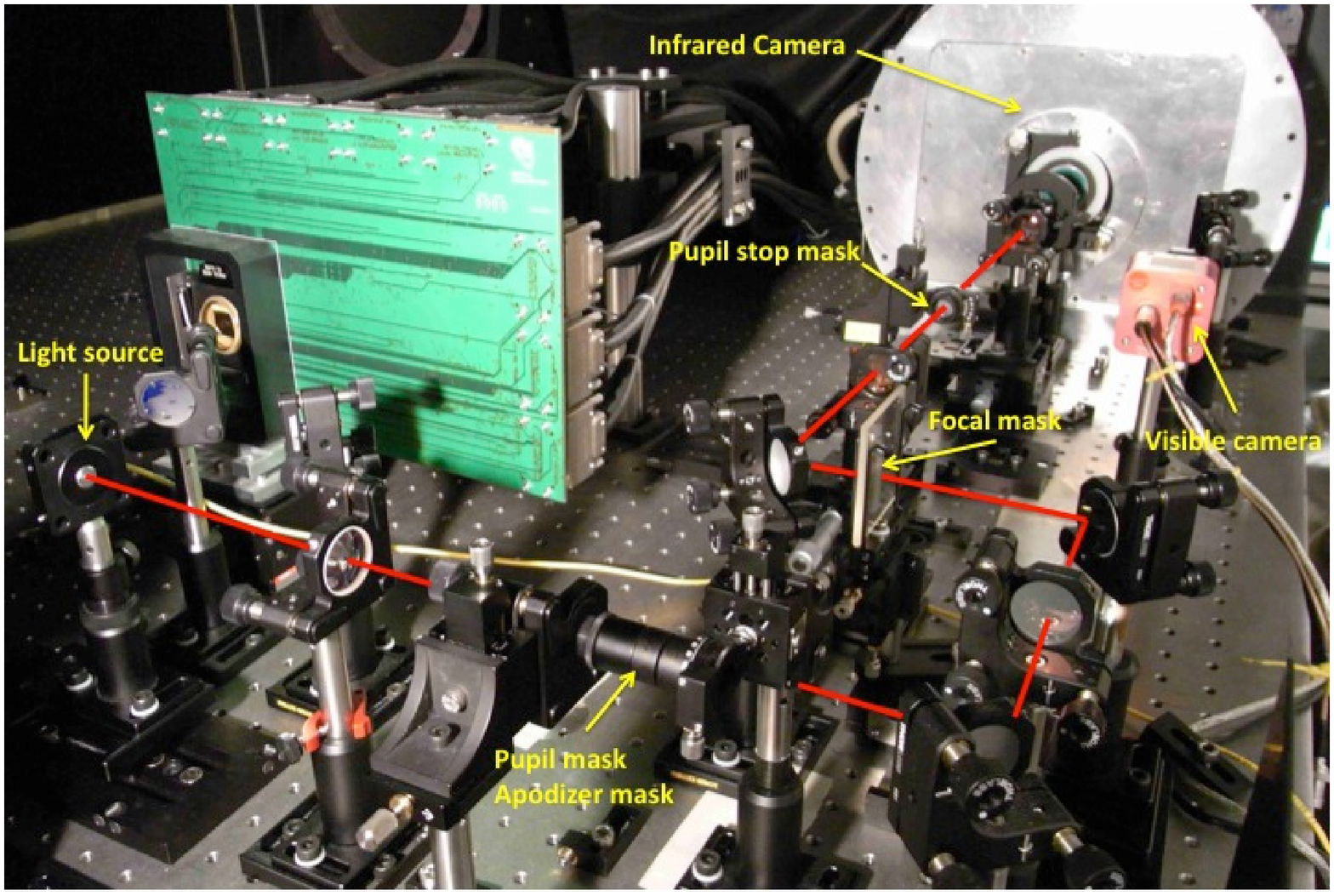}
\caption{From the top to the bottom: details of the near-IR arm of HOT (science channel), and its configuration when used as an independent coronagraphic testbench (a folding mirror is replaced by an entrance fiber).}
\label{BENCH2}
\end{figure}

It is fundamental to evaluate the intrinsic capabilities and limitations of each coronagraphic concepts on a dedicated optical testbench before testing the coronagraphs on HOT under dynamical seeing and XAO correction.
This is essential to appropriately interpret experimental contrast results obtained on HOT afterwards, i.e., to distinguish the impact of the aberrations left uncorrected by the XAO system from the limitations introduced by the coronagraph prototypes themselves. In practice this is possible by using the science arm of HOT (near-IR optical path) as an independent coronagraphic optical setup. 
The relevant modifications made on the science arm of HOT are described in Fig. \ref{BENCH2} where one can compare the standard configuration (top) to that of the coronagraphic testbench configuration (bottom). The main difference comes from the fact that a flat mirror used to fold the beam after the DM is replaced by an entrance fiber, which is used as a star source for the optical setup (6.6 $\mu$m diameter).
The coronagraphic testbench uses a VLT-pupil mask made in a laser-cut, stainless-steel sheet. 
The dimensions and positioning of the VLT-pupil in both the coronagraphic testbench and HOT are alike.

The Strehl ratio of the coronagraphic testbench was evaluated in the $H$-band to be 93$\%$ $\pm$ 1$\%$ by measuring the peak-intensity ratio of the experimental PSF to that of the theoretical PSF normalized to the total intensity.
The theoretical PSF is created through two different methods, both leading to the same Strehl ratio:
(1/) by performing the forward fast Fourier transform (FFT) of the autocorrelation of an oversampled and uniformly illuminated entrance pupil image from our telescope pupil mask, 
(2/) by performing the FFT of a simulated aperture function with radius determined from the experimental PSF on the basis of photometric criteria.

As high-frequency wavefront components impose an important contrast limitation, we determined the wavefront error of the optical components prior to the pupil-stop in the coronagraphic testbench to an overall total amount of $\sim\lambda/67$ rms at 1.6 microns, i.e., 24 nm rms \citep{martinez09c}. 

All the parameters described below are identical when operating with the HOT or the coronagraphic testbench: the VLT-pupil in the first pupil-plane of the IR arm has 3 mm diameter, a central obscuration of  0.47 mm, and 4 spider-vanes with 15 $\mu$m thickness. The coronagraphic focal masks are installed at an F/48.4 beam. The pupil is re-imaged in the Lyot plane where the pupil-stop is installed, and appears with similar dimensions as in the first pupil-plane of the IR arm. All re-imaging optics are made with $\lambda$/10 achromatic IR doublets. 
Therefore, all coronagraphs can be tested similarly on HOT and on the coronagraphic testbench.

\subsection{Definition of metrics}
Several metrics can be used to quantify the capability of a coronagraph to suppress the on-axis starlight: 
\begin{itemize}
\item The \textit{rejection rate ($\tau$)}: ratio of total intensity of the direct image to that of the coronagraphic image.
\item The \textit{peak rejection rate ($\tau_0$)}: ratio of the maximum intensity of the direct image to that of the coronagraphic image.
\item The \textit{contrast ($\mathscr{C}$)}: ratio of the coronagraphic image at a given angular separation to that of the maximum of the direct image to the intensity, azimuthally averaged.
\end{itemize}
\noindent In addition, as contrast evaluation is no longer suitable on post-processed coronagraphic images, the azimuthal standard deviation -- quantifying the ability to pick out a companion at a given angular distance -- will be applied in such situation.
The azimuthal standard deviation is measured in a ring of width $\lambda/D$ on the post-processed image, i.e., it includes all pixels within one resolution element. Although the detectability is commonly estimated at the level of 5$\sigma$ on real data \citep[e.g.,][]{hinkley07} and since we are not sensitive to confidence level threshold with our data, we quantify the detectability at the level of 1$\sigma$.
We adopt the azimuthal contrast and standard deviation as primary metrics although we note they are  conservative estimates, and that other criteria adapted to the case of high-contrast images have been discussed by \citet{marois08}.

\subsection{Pupil stops optimization}
The design of the coronagraph pupil-stops (Fig. \ref{STOPS}) emphasizes the minimization of diffraction effects (concentration of light around the edges, spider vane structures, and central obscuration) while preserving the throughput. All pupil-stops have been optimized following the metric presented in \citet{boccaletti04b}, which attempts to maximize the ratio of the off-axis PSF throughput to the stellar diffraction residuals with the use of homothetic stop-patterns (same shape as the VLT-pupil but under/over sized). 
This optimization is performed at the operating wavelength in a wavefront error-free condition (i.e., Strehl equal to 100$\%$).
The performance of the coronagraph combined with the optimized pupil stop is afterwards compared with intrinsic or external limitations affecting the coronagraph efficiency (e.g., stellar size, chromaticism...). Based on the analysis of this budget error and accounting for alignment considerations the optimized pupil stop is retained, relaxed, or further optimized. 

\subsection{Chromaticity considerations}
A smooth and flat chromaticity response of the coronagraph is paramount for exoplanet detection as broadband observations are required for spectroscopy. For instance, most of the next generation of high-contrast imaging instruments will use integral field spectroscopy for data collection. Analyzing coronagraphic performance in a large range of small bandwidths would therefore be extremely relevant. But it requires a dedicated optimization of each coronagraph that can only be carried out in correlation with specific observation and data reduction strategies, which is beyond the scope of this study and beyond the capabilities of our experimental setup. Therefore the chromatic propagation (contrast vs. wavelength) of each coronagraph cannot be addressed in this study, although it is certainly a relevant comparative criterion.

\section{Coronagraph prototypes}
\label{PROTO}
In this section, we provide details of our four coronagraph prototypes and their performance. 
We first examine the impact of several error sources on their respective efficiency by probing the effect of manufacturing, wavefront error, chromaticity, and the source diameter. A performance characterization realized on the coronagraphic testbench will be presented and discussed afterwards.
This first study is a prerequisite for proper analysis and interpretation of  the results obtained on HOT under dynamical seeing and XAO correction, i.e., to identify the limitations occurring in the coronagraphic images.

\subsection{Four-quadrants phase mask}
The four-quadrant phase mask (FQPM) is a phase coronagraph proposed by \citet{rouan00}. The mask divides the focal plane into four areas, two of which being $\pi$ phase-shifted.  As a result, a destructive interference occurs in the relayed pupil, where the on-axis starlight is rejected and filtered by an appropriate pupil-stop. This concept has been studied from a theoretical point of view \citep{rouan00, riaud01}, in the laboratory \citep{riaud03}, and has provided on-sky results \citep[e.g.,][]{Boccaletti04, gratadour06, riaud06, Boccaletti08}.

Our prototype was manufactured by GEPI (Galaxies Etoiles Physique et Instrumentation) in collaboration with LESIA (Laboratoire d\textquoteright Etudes Spatiales et d\textquoteright Instrumentation en Astrophysique) both from the Paris Observatory. The prototype is a monochromatic device, i.e., the $\pi$ phase shift is obtained at a single wavelength. 
Achromatic FQPMs can be made using half-waves plates \citep[][SPHERE coronagraph]{mawet06}, AGPM \citep[annular groove phase mask,][]{mawet05}, or OVC \citep[optical vortex coronagraph,][]{mawet09}. 
We also note that achromatic implementation of the FQPM exist \citep{baudoz07}.

The FQPM is manufactured by engraving of two opposite quadrants on an optical medium. The substrate is made in INFRASIL  301 with 16 mm diameter and 3 mm thickness ($\pm$ 0.1 mm).
The optical quality for the two faces is $\lambda$/20 peak-to-valley at 633 nm.
The thickness (e) and optical index (n) of the layer has been defined to provide a $\pi$ phase shift at the operating wavelength $\lambda_0$  = 1.65 $\mu$m ($H$-band) following:
\begin{equation}
\Phi = \frac{2\pi (n - 1) e}{\lambda_0}
\label{eq4Q}
\end{equation}  
As the thickness of the FQPM step (e) directly defines the optimal wavelength at which the attenuation is best, an error between specified and manufactured step reduces the of the FQPM efficiency. 
Our specification was  e = 1.89 $\mu$m $\pm$ 3 $\%$.
A dedicated visible spectroscopic bench was used at LESIA to measure the thickness of the FQPM step \citep{riaud03, Boccaletti04} by measuring the wavelength corresponding to the optimal nulling in the near-IR.
In practice this facility enables to record low resolution spectra with a source centered on the FQPM (coronagraphic spectrum) and with the source out of the FQPM (direct spectrum). The ratio of these two spectra allows one to identify different coronagraphic minima that correspond to a phase difference between the quadrant of $\Delta \Phi = k \times \pi$ (with k = 1, 3, 5, 7 and so on). 
From results of a given identified order in the visible (odd value of k), it is straightforward to derive from these data and from the known optical index of the material, the operating wavelength of the FQPM ($\lambda_0$) at $k=1$.
Equation \ref{eq4Q} can then be used to extract the thickness (e).
While a precision of less than 3$\%$ was required,  a depth accuracy of 0.2$\%$ was finally obtained after several runs.

Ideally the transition between the four quadrants must be infinitely sharp. Departure from this ideal case decreases the capability of the real device. Microscopic inspection of the manufactured FQPM have shown that the transition quality is faster than 1 $\mu$m (2 $\mu$m peak-to-peak transition). The accuracy on the quadrant orthogonality has been measured to be $\le$ 0.8 arcmin. 

The manufactured FQPM is sensitive to several parameters as discussed previously (Table \ref{FQPMtable}). Specific care has been taken when defining the pupil-stop (Fig. \ref{STOPS}, left) in order not to allow higher rejection rate than the limit imposed by inherent parameters from the prototype. 
It is indeed useless to define an aggressive pupil-stop in diffraction-limited regime but rather convenient to save throughput through a relaxed optimization when external error sources, e.g.,  the stellar leakage, chromaticity, will first set the limitation in the efficiency of the coronagraph. Table \ref{FQPMtable} quantifies all manufacturing defects and error sources that impact the FQPM efficiency, and will be discussed in the next subsection. These data were used during the optimization process. The FQPM pupil-stop mimics the VLT pupil mask (see Fig. \ref{STOPS}, first pattern from the left) with a spider-vane thickness larger by a factor 5.5 (82.5 $\mu$m $\pm$ 4 $\mu$m), an outer diameter smaller by a factor 0.90$\times \Phi$ (2.70 mm $\pm$ 0.002 mm), and a central obscuration equal to $\sim$0.3$ \times \Phi$ (0.90 mm $\pm$ 0.002 mm). The pupil stop transmission is 70$\%$. 

 
\begin{figure}
\centering
\includegraphics[width=8.0cm]{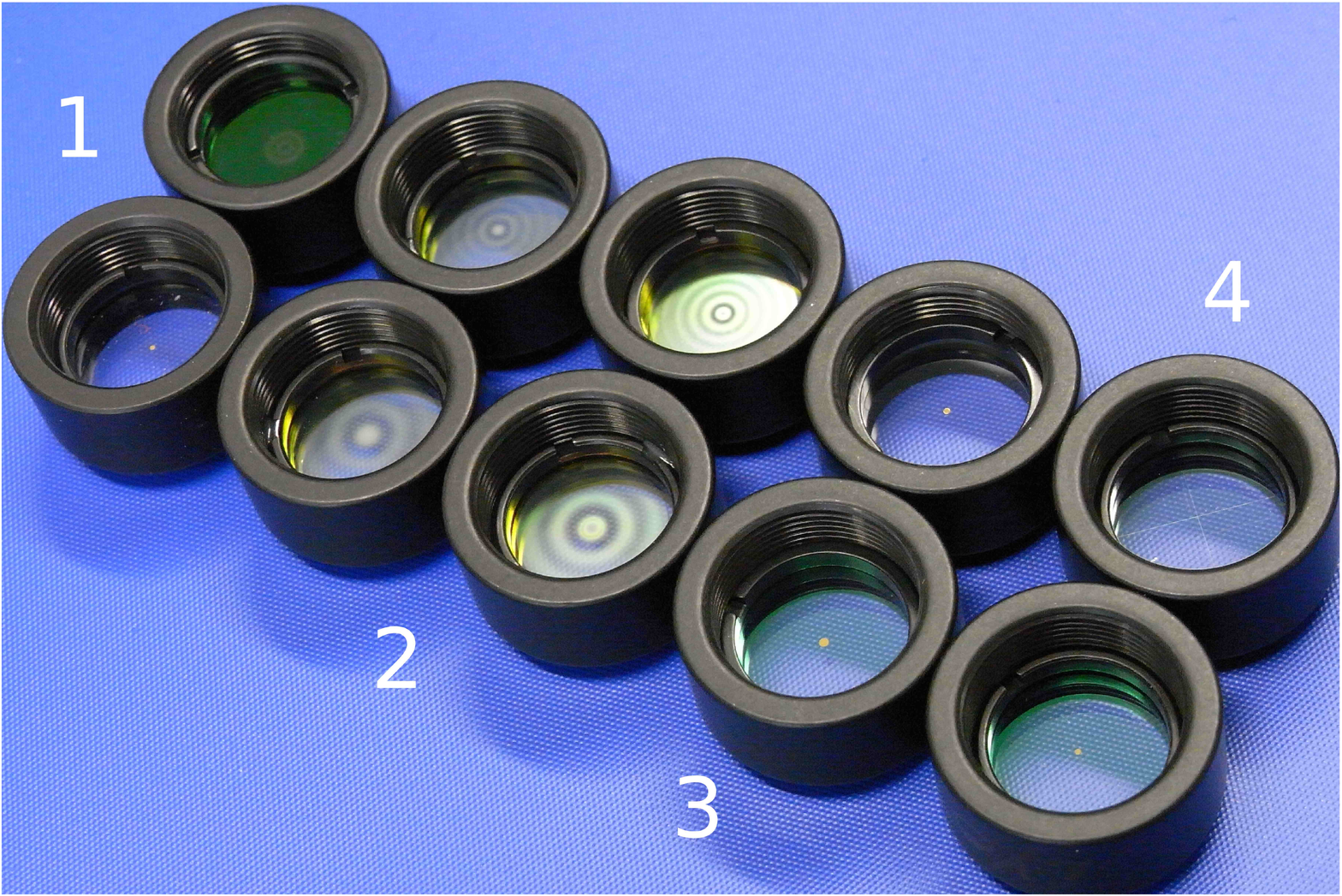}
\caption{Various coronagraphic devices manufactured and tested on HOT: APLC (1), BLCs (2), LCs (3), and FQPM (4).}
\label{PROTOS}
\end{figure} 
\begin{table}
\centering
\begin{tabular}{l|r}
\hline \hline 
Parameter & Achievable rejection ($\tau$) \\
\hline \hline 
Step thickness (0.2$\%$) & 120668 \\
Quadrant transition (1 $\mu$m) & 1890 \\
Chromaticity (R=1.4$\%$) & 23830 \\
Chromaticity (R=24$\%$) & 121 \\
Wavefront error ($\lambda/67$ rms at 1.64 $\mu$m) & 428 \\ 
Source diameter (6.6 $\mu$m, i.e., 0.08$\lambda/D$) & 1350 \\
Central obscuration ($15\%$) &  680 \\
\hline
\end{tabular}
\caption{Manufacturing defects and error sources impact on the total rejection rate of the FQPM.}
\label{FQPMtable}
\end{table}
 
\subsubsection{Coronagraphic capabilites} 
Table \ref{FQPMtable} presents the manufacturing defects of the prototype and several error sources that reduced the FQPM efficiency.
The formula quantifying the impact of these particular error sources are not detailed here since they are available in the literature \citep[e.g.,][]{riaud01, riaud03}. 
Several aspects are considered such as manufacturing defects, chromaticity (depending on the filter bandpass we use in the experiment), stellar leakage (the FQPM is very efficient to observe at close angular separations of the star but highly sensitive to the stellar angular size), high-frequency wavefront errors of the coronagraphic testbench, and the central obscuration of the pupil (VLT-like, performance of the FQPM degrades with the central obscuration size). As all these aspects can be quantified it is straightforward to estimate the best performance that can be reached with our prototype.  Assuming that all these independent errors are added quadratically, the expected global rejection factor ($\tau$) is found to be $334$ and $114$ for 1.4$\%$ and 24$\%$ spectral bandwidth, respectively.   

The FQPM coronagraphic characterization is presented in Table \ref{COROtable}. Both narrow and broadband runs have led to results compatible with the limitations previously described ($\tau = 278$ and $77$ for $\Delta\lambda/\lambda=1.4\%$ and 24$\%$, respectively, while the expected nulling for similar bandwidth were 334 and 114). 
The difference between experimental and expected values may come from the uncertainty of the filter bandpass (more likely in the case of the polychromatic test, i.e. the actual filter spectral transmission is not taken into account), or typical errors  that arise in a coronagraphic system and neglected here (offset pointing, alignment issue, pupil shear and rotation...). 
With these considerations in mind, the agreement between experiment and expectation is fairly good.

The FQPM contrast evolves from $\sim10^{-4}$ at short angular separations ($\theta < 3\lambda/D$) to 3$\times10^{-6}$ in the halo ($\theta = 20\lambda/D$). Being a monochromatic device, the FQPM is sensitive to the filter bandpass, where short angular distances are mostly affected. At IWA ($\sim$1$\lambda/D$), the contrast delivered is 4$\times10^{-3}$ and 9.5$\times10^{-4}$ in monochromatic and polychromatic images respectively.
 
\subsection{Lyot coronagraphs}
A set of 3 Lyot coronagraphs (LCs) with different mask diameters have been manufactured using wet-etch lithography on BK7 glass by GEPI.
They are made using Cr deposition (+Au) to reach an OD (optical density) of 6.0 at 1.65 $\mu$m. The diameters are 360, 390, and 600 $\mu$m (i.e., 4.5, 4.9, 7.5$\lambda/D$ respectively, defined hereafter as LC(1), LC(2), and LC(3) respectively).
The diameter accuracy is $\sim$1 $\mu$m. Inspection with a microscope has been carried out and confirmed the perfectly circular shape. 

\begin{figure*}
\includegraphics[width=4.cm]{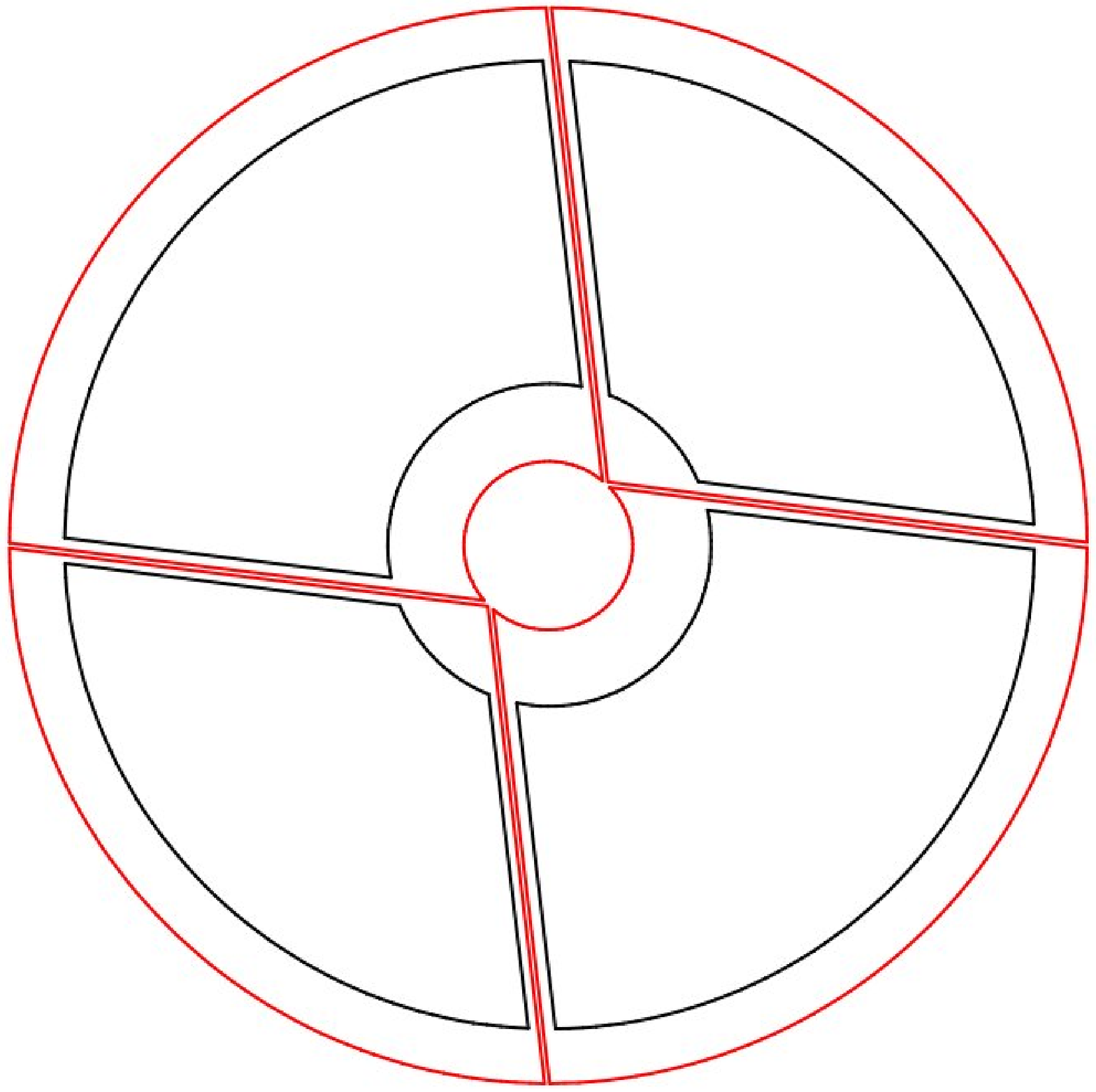}
\includegraphics[width=4.cm]{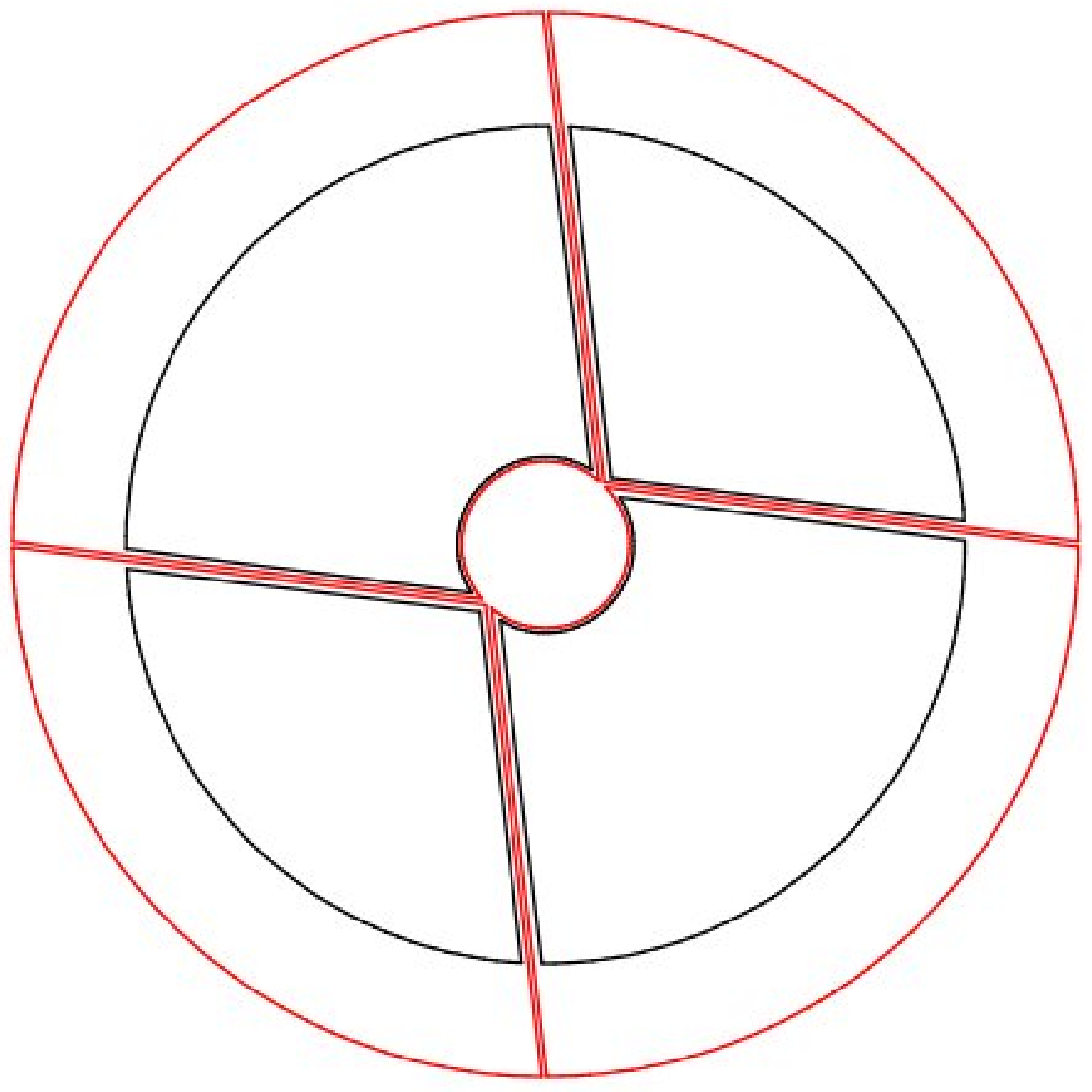}
\includegraphics[width=4.cm]{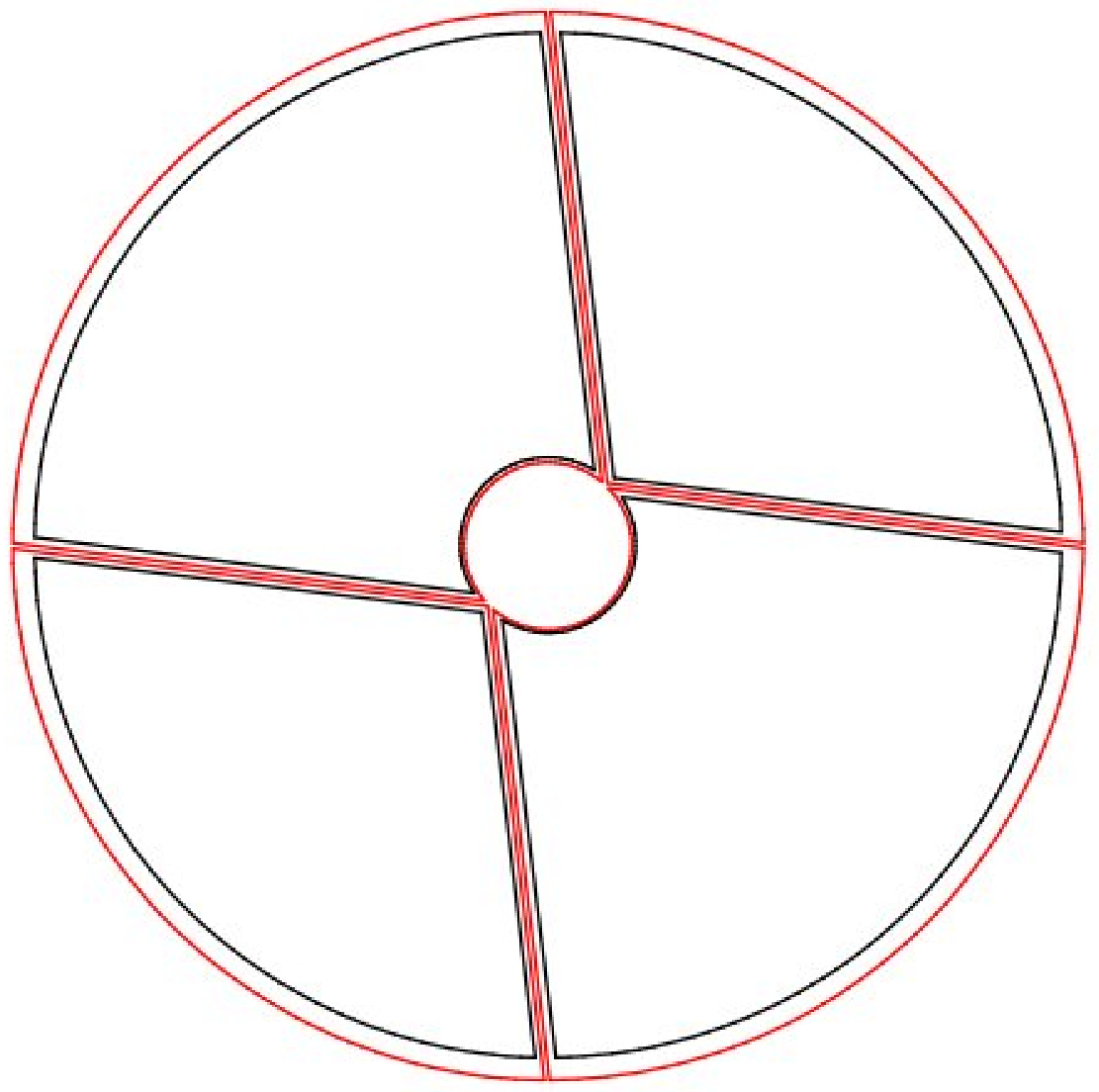}
\includegraphics[width=4.cm]{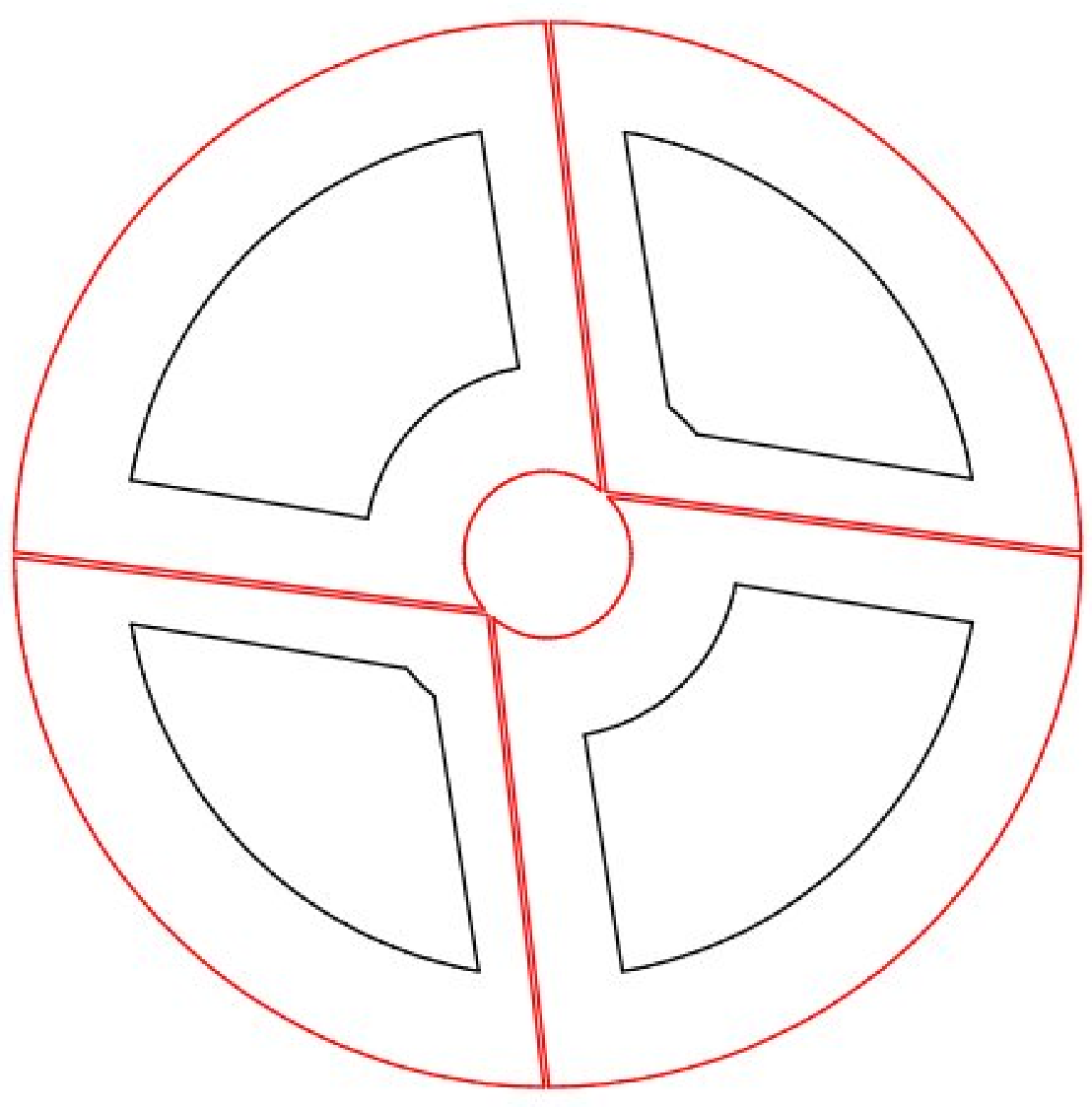}
\caption{From the left to the right: as specified pupil stops (black) superimposed to the VLT-pupil (red) for the FQPM, LC, APLC, and BLC.}
\label{STOPS}
\end{figure*}

The LC pupil-stop throughput is $\sim60\%$ and has been optimized for the smallest Lyot mask diameter (4.5$\lambda/D$), being usable for the two others, avoiding multiple pupil-stops for the LCs.
As the Lyot mask diameter increases, the diffracted light in the pupil stop is more localized and concentrated.
The LC pupil-stop mimics the VLT pupil mask (see Fig. \ref{STOPS}, second pattern from the left) with a spider-vane thickness larger by a factor 4 (60 $\mu$m $\pm$ 4 $\mu$m), an outer diameter smaller by a factor 0.78$\times \Phi$ (2.36 mm $\pm$ 0.002 mm), and a central obscuration equal to $\sim$0.17$\times \Phi$ (0.50 mm $\pm$ 0.002 mm). 

\subsubsection{Coronagraphic capabilities} 
As the manufacturing aspects of the LCs do not present any particular issues or difficulties (i.e., the LC is the easiest concept to manufacture in practice) no error budget has been set. By contrast to the FQPM, stellar leakage, central obscuration, manufacturing defects, or chromaticity are not strongly limiting factors. With the as-manufactured pupil-stop and focal masks, simulations assuming ideal conditions have shown that the expected total rejection rate ($\tau$) in a monochromatic case is 250, 260, and 630 for the 4.5, 4.9, 7.5$\lambda/D$ (LC(1) to LC(3)) masks respectively.

The characterization results are presented in Table \ref{COROtable}. The monochromatic run ($\Delta \lambda/\lambda=1.4\%$) nicely fulfilled expectations (total rejection rate predicated from simulations). The measured contrast evolves from $\sim10^{-4}$ at close angular separations ($\theta \sim 3\lambda/D$) to 4 to 6$\times10^{-6}$ at best at farther separations ($\theta = 20\lambda/D$). Apart from the peak attenuation, no real gain is observable by using larger mask configurations (e.g., LC(3)).  

All LCs demonstrate almost similar performance from $\Delta \lambda/\lambda$=1.4$\%$ to $\Delta \lambda/\lambda$=24$\%$, and therefore do not present strong chromaticity dependence. 

\begin{figure*}
\includegraphics[width=8.5cm]{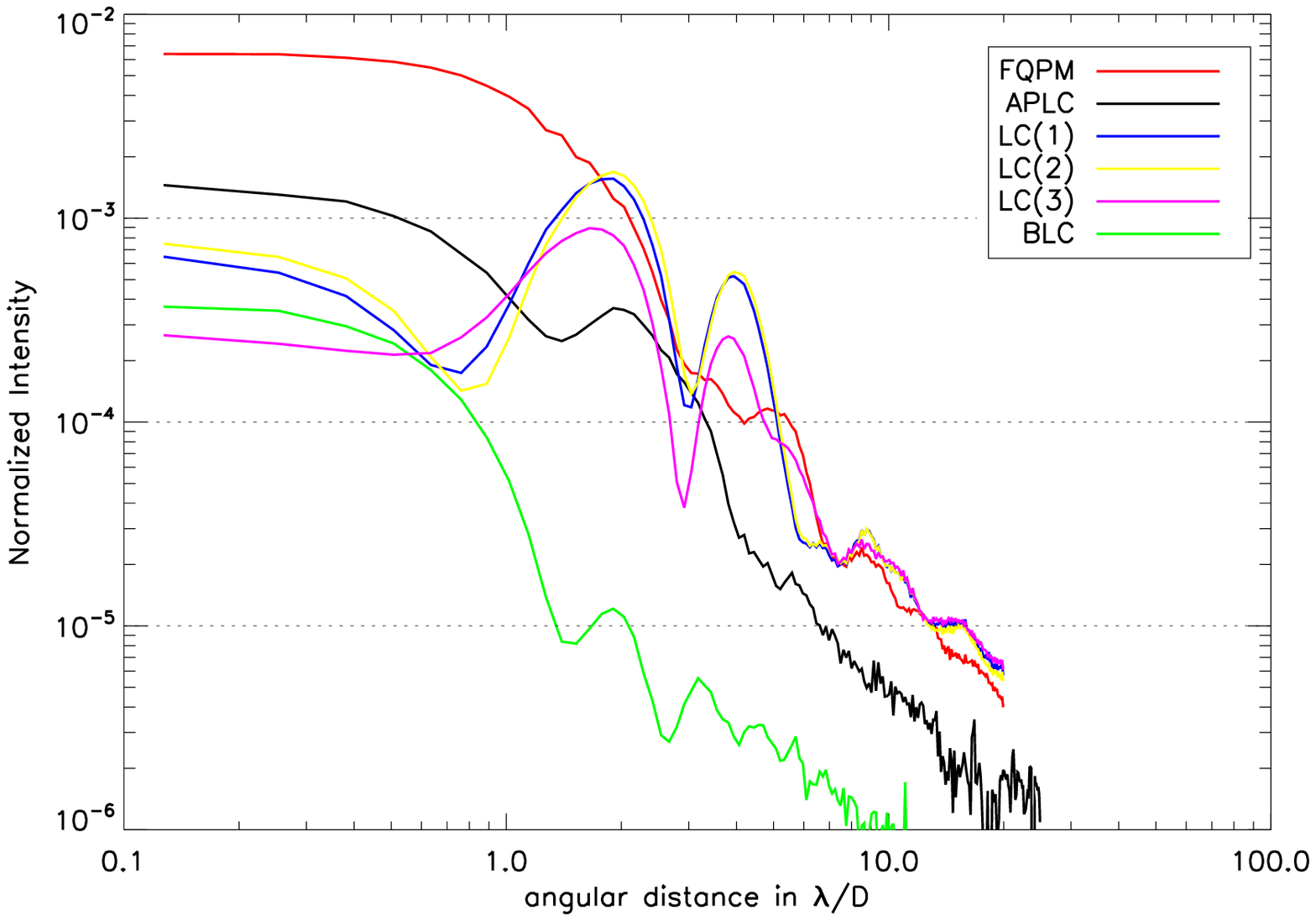}
\includegraphics[width=8.5cm]{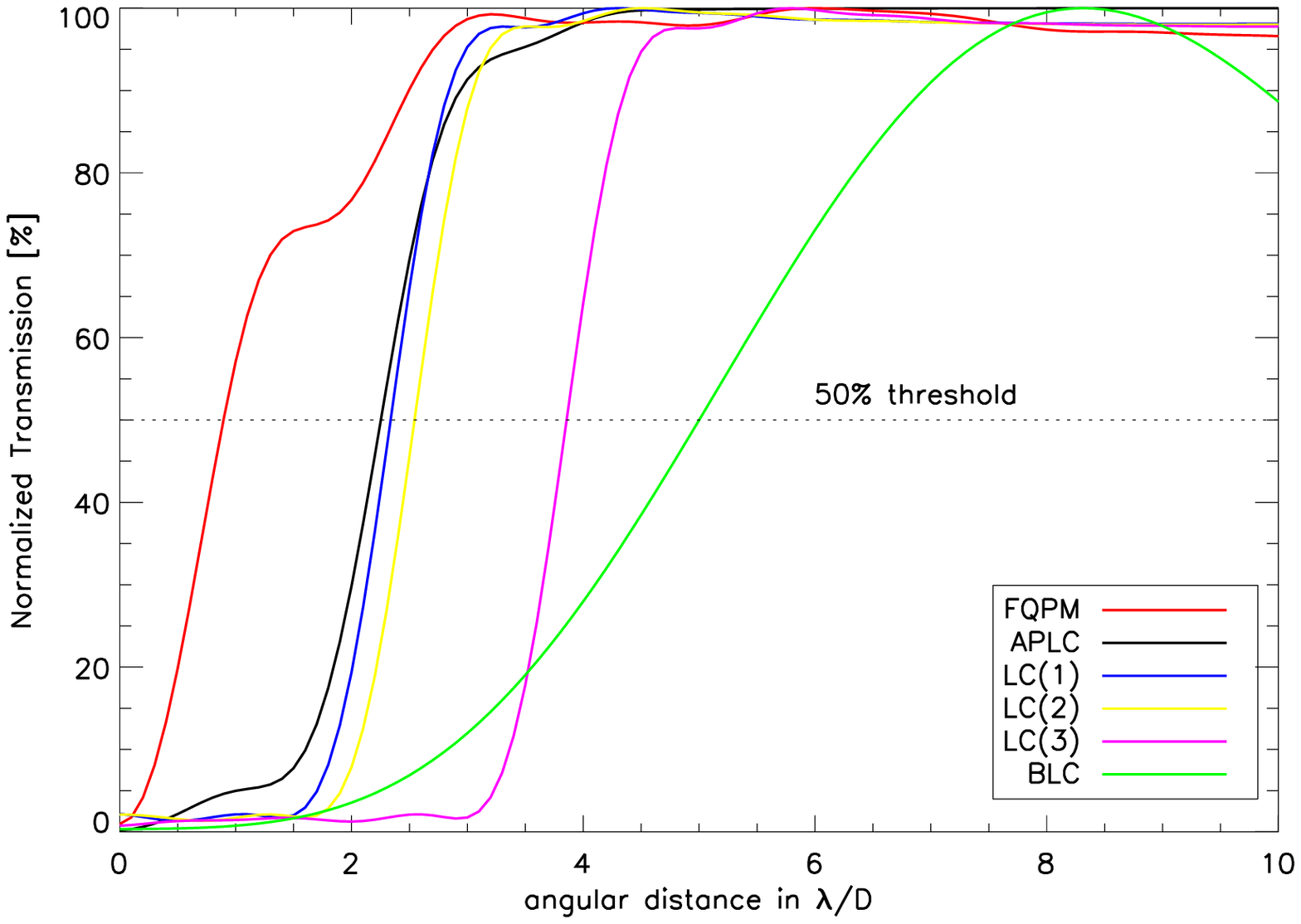}
\caption{Left: Raw coronagraphic contrast profiles azimuthally averaged ($\Delta \lambda/\lambda = 24\%$). Right: Radial transmission of each coronagraphs.}
\label{Results0}
\end{figure*} 
\begin{table}
\centering
\begin{tabular}{l|l|l|l|l|l|l|}
\hline \hline 
 & BW ($\%$) & $\tau$ & $\tau_0$ & $\mathscr{C}_{3\lambda/D}$ & $\mathscr{C}_{12\lambda/D}$ & $\mathscr{C}_{20\lambda/D}$ \\ 
\hline \hline 
 \multicolumn{7}{c}{Raw images} \\
 \hline \hline
FQPM & 1.4 & 278 & 760 & 7.9 $10^{-5}$ &  9.4 $10^{-6}$ & 2.9 $10^{-6}$\\ 
            &  24 & 77 & 156 & 1.9 $10^{-4}$ & 1.2 $10^{-5}$ & 3.9 $10^{-6}$\\ 
\hline
LC (1)   & 1.4 & 239  &   586 & 1.0 $10^{-4}$ & 9.9 $10^{-6}$ & 4.9 $10^{-6}$\\ 
            &  24 &  247 & 640 & 1.2 $10^{-4}$ & 1.2 $10^{-5}$ & 5.7 $10^{-6}$\\
 \hline
LC (2)   & 1.4 & 253  &   606 & 9.5 $10^{-5}$ &  9.0 $10^{-6}$ & 3.8 $10^{-6}$\\ 
            &  24 &  231 & 593 & 1.7 $10^{-4}$ & 1.3 $10^{-5}$ & 5.3 $10^{-6}$\\ 
 \hline
LC (3)    & 1.4 & 586 &1477 & -- & 8.5 $10^{-6}$ & 3.8 $10^{-6}$\\ 
            &  24 & 408 & 1120 &-- & 1.2 $10^{-5}$ & 6.0 $10^{-6}$\\ 
 \hline
APLC & 1.4 & 489 &    627 & 5.0 $10^{-5}$ & 2.3 $10^{-6}$ & 1.2 $10^{-6}$ \\
            &  24 &  355 & 674 & 1.5 $10^{-4}$ & 3.5 $10^{-6}$ & 1.8 $10^{-6}$ \\
 \hline
BLC    & 1.4 & -- & -- & -- & -- & -- \\
            &  24 & 2410 & 2554 & -- & 5.6 $10^{-7}$ & 2.6 $10^{-7}$\\
\hline
\end{tabular}
\caption{Performance of each coronagraph concepts obtained when tested on the coronagraphic testbench (turbulence-free).}
\label{COROtable}
\end{table}
\begin{table}
\centering
\begin{tabular}{lcc}
\hline \hline 
Concept & IWA ($\lambda/D$) & Throughput ($\%$) \\
\hline \hline 
FQPM & 1.0 & 70 \\
LC(1) &   2.3     &    60    \\
LC(2) &   2.5     &    60    \\
LC(3) &    3.8    &    60    \\
APLC & 2.3 &  45 \\
BLC &  5.0  &   42  \\
\hline
\end{tabular}
\caption{Coronagraphs main characteristics.}
\label{Resumtable}
\end{table}
\subsection{Apodized-pupil Lyot coronagraph}
\subsubsection{Prototype details}
The apodized-pupil Lyot coronagraph \citep[APLC,][]{aime02, soummer03} combines pupil apodization with hard-edge focal plane mask. 
Our APLC is a 4.5$\lambda/D$ mask diameter configuration coupled with its corresponding optimized bagel-shaped pupil apodizer. 
This APLC configuration is specifically adapted for obstructed entrance apertures \citep{soummer05, martinez07, soummer09}.   
The prototype has been detailed in a previous paper \citep{martinez09a}. Briefly, the 4.5 $\lambda/D$ hard-edge opaque Lyot mask was fabricated by GEPI at the Paris Observatory (360 $\mu$m $\pm$ 1 $\mu$m in diameter, OD = 6.0 at 1.65 $\mu$m using two metallic layers of Chrome (20 nm) and Gold (200 nm)), while the apodizer has been manufactured by Precision Optical Imaging in Rochester, New York, and is made with binary metal pixels \citep{martinez09a, martinez09b, martinez10b}.The profile accuracy is of about 3$\%$ and the transmission of the apodizer is $\sim$ 50$\%$.  \\ 
The APLC pupil stop  (Fig. \ref{STOPS}, third pattern from the left) mimics the VLT pupil mask with a spider-vane thickness larger by a factor 4 (60 $\mu$m $\pm$ 4 $\mu$m), an outer diameter smaller by a factor 0.96$\times \Phi$ (2.88 mm $\pm$ 0.002 mm), and a central obscuration equal to 0.16$\times \Phi$ (0.49 mm $\pm$ 0.002 mm). The pupil stop throughput is about 92$\%$.

\subsubsection{Coronagraphic capabilities} 
In the coronagraphic testbench we installed the entrance-pupil mask (VLT-like pupil) and the apodizer mask in the same collimated beam because of the lack of space. Therefore, the apodizer cannot be in practice in the same plane as the VLT pupil mask.
To minimize this imaging issue, the apodizer was placed inside a rotating adjustable-length lens tube that allows translation motion along the collimated beam, and was adjusted at $\sim$3.5 mm from the pupil mask.
The APLC performance is presented in Table \ref{COROtable}, where an order of magnitude discrepancy (mostly in the halo) was found between theory and measured data \citep{martinez09a}. Some error sources impact the APLC performance, such as the apodizer profile error ($\sim3\%$), or the defocus between the apodizer and the pupil plane ($\sim3.5 mm$) although the depth of focus at the pupil is $\sim$7 mm.
Because the quality of the optical setup also plays a role it is difficult to estimate the respective impact of each aspects. However, it is known that the apodizer positioning error in focus mainly affects the halo level, while the apodizer profile error affects both peak level and contrast in the halo. In addition, typical defects that occur at the level of the focal mask (pointing error, defocus) and at the pupil-stop (shear and rotation) impact the APLC performance as well. All these aspects acting together may explain the discrepancy between data and theory.  
The contrast evolves from $\sim10^{-4}$ at close angular separations ($\theta > 3\lambda/D$) to $\sim10^{-6}$ at best at farther separations ($\theta = 20\lambda/D$).
The impact of chromatism is slightly observable at small angular separations (less than 4$\lambda/D$), but the halo is found to be achromatic. 
In addition, we note that dedicated studies aimed to mitigate the chromatic response of the APLC \citep[e.g.,][]{martinez07, soummer11}.

\subsection{Band-Limited coronagraph}
\subsubsection{Prototype details}
The Band-Limited coronagraph \citep[BLC][]{kuchner02} is an improvement of the LC concept through the use of a specific design of the amplitude focal plane mask. The mask has a power spectrum with power in a limited range of frequencies insuring that the mask is designed to both remove starlight and the diffraction effects caused by the removal of the light. By principle the BLC is achromatic. 
Our prototype is based on a band-limited function proposed by \citet{kuchner02}:
\begin{equation}
M(r) = N \left( 1 - sinc\left(\frac{\epsilon r D}{\lambda}\right) \right)
\label{function}
\end{equation}

\noindent where $\lambda$ is the wavelength of the application, $r$ the radial coordinates in the image plane, D the telescope primary diameter, $\epsilon$ the bandwidth which rules the inner-working angle of the coronagraph (IWA hereafter), and finally N is a constant of normalization insuring that $0 \leqq M(r) \leqq 1$.

Our prototype corresponds to $\epsilon=0.17$ (i.e. IWA = 5$\lambda/D$) and was manufactured by Precision Optical imaging (Rochester, New York) with binary metal pixels \citep{martinez09c}.
The mask was designed for 1.64 $\mu$m, and fabricated using wet-etch contact lithography of an Aluminum layer ($OD=8+$, $e=2000\dot{\mathrm{A}}$) deposited on a BK7 substrate ($\lambda/10$ peak-to-valley, 0.5 inch diameter). Antireflection coating in $H$-band has been applied on each face.
The BLC uses 5$\mu$m pixels.  Profile accuracy is of about $5\%$ of the specification, where the error is mostly localized in the outer part of the mask (high-transmission part), as the center part (for the low-transmissions) is highly accurate \citep{martinez09c}. The error in the outer part originates in a calibration issue of the process, that was later corrected with new prototypes demonstrating a profile error below 1$\%$.

The BLC pupil stop optimized for HOT  (i.e., designed for the VLT pupil, Fig. \ref{STOPS}, right) has been manufactured with a spider-vane thickness larger than the entrance pupil by a factor 27 (0.4 mm $\pm$ 4 $\mu$m), an outer diameter that is smaller by a factor 0.80$\times \Phi$ (2.40 mm $\pm$ 0.002 mm), and a central obscuration that is equal to 0.35$\times \Phi$ (1.05 mm $\pm$ 0.002 mm). The pupil stop throughput is about 42$\%$. 
By principle, the BLC can accommodate arbitrary telescope apertures with proper pupil-stop optimization \citep{kuchner02}, although this can usually impose aggressive reduction of the pupil when dealing with sophisticated telescope apertures \citep[e.g. ELTs,][]{martinez08}, as in our situation.

The BLC characterization on the coronagraphic testbench is presented in Table \ref{COROtable} and has been obtained with a clear aperture and a 70$\%$ throughput pupil-stop as the pupil stop optimized for the VLT pupil was not available during the run carried out on the coronagraphic testbench. However, a new prototype of this BLC was recently tested on the coronagraphic testbench with the VLT-pupil and its corresponding pupil-stop yielding to similar contrast level in the halo as the one discussed in the next subsection, while the peak and total rejection rate were improved by more than a factor of 2.  

\subsubsection{Coronagraphic capabilities} 
The BLC has demonstrated impressive performance on the coronagraphic testbench (Table \ref{COROtable}), contrasts evolve from $\sim3\times10^{-5}$ at IWA to $\sim3\times10^{-8}$ at $20\lambda/D$, while the peak rejection is 2554 (i.e. more than four times the peak rejection of most of the other concepts). More details on the BLC manufacturing and characterization can be found in \citep{martinez09c}, where achromaticity of the concept has been confirmed. 

\section{Results under dynamical seeing and XAO}
While in the previous section we discussed the characterization of our prototypes on the coronagraphic testbench,  the present section is dedicated to the test of all the coronagraphs on HOT under dynamical seeing and XAO correction. 

In Table \ref{Resumtable}, we recall the IWA and throughput for each concept. In addition, Fig. \ref{Results0} (right) compares the radial transmission of each coronagraph, which provides the transmission of an off-axis companion at close and farther radiis. Figure \ref{Results0} (right) has been obtained by simulation given the shape of the entrance VLT-pupil and pupil stop, and each coronagraphic mask used in the experiment. 
One should note that contrast evaluation presented in this paper does not account for the radial transmission of each coronagraphic mask, which is especially important at close radii from the axis. Indeed from Fig. \ref{Results0} (right) one can observe that the FQPM allows access to very short angular separations that cannot be observed by others.

\subsection{Experimental conditions}
During the experiment, the XAO system was operating with the Shack-Hartmann wavefront sensor (SHWFS).
The dynamical turbulence is generated using two phase screens in reflection rotating independently from each other, and specified to reproduce a turbulence of 0.5$\arcsec$ seeing.
Measurements with an HASO 64 Shack-Hartmann sensor have been carried out to verify the phase screens parameters.  
The power spectrum of the reconstructed wavefront was compared to the theoretical and specified Von Karman spectrum (\textit{$L_0$} = 25m, and \textit{$r_0$} = 12cm) with good agreement.
The phase screens are low-order aberration reduced to avoid saturation of the DM (limited actuator stroke of about two microns).

The static aberrations, corresponding to common-path wavefront errors, have been reduced from $\sim$200 nm rms to 50 nm rms using a bimorph-DM in open-loop, i.e., by applying a corrective voltage pattern to the initial voltage pattern that setup the mirror. The SHWFS non-common path wavefront errors have been measured using a reference fiber at the entrance of the SHWFS, and 
slope offsets have been applied to the DM not to take into account these aberrations in the correction. 
Additionally, some slope offsets have been applied to correct for non-common path errors from the near-IR optical path on the basis of the PSF image quality.
Further details can be found in \citet{vernet06, carpentier08}. 

The SHWFS closed-loop runs at 80Hz using 600 modes for the modal reconstruction on a 8-meter pupil. In 0.5$\arcsec$ seeing with 1.3m/s wind speed, the system delivers a 90$\%$ Strehl ratio in $H$-band at high-flux (star magnitude of 5). 
The DM actuator pitch being equal to 340 $\mu$m, the AO cut-off frequency is localized in the field at an angular separation of 15$\lambda/D$, i.e., 0.6$\arcsec$ for a 8-meter telescope in $H$-band. 

The experiment was carried out with a series of 3-second short exposure images averaged over 3 minutes, and neutral density filters were applied on non-coronagraphic images only. 
Dark frames were obtained by switching off the artificial star source.
The data reduction corrects bad pixels, background, and scales images by the exposure time and optical density.

While the contrast profiles are evaluated on raw coronagraphic images, all coronagraphic images were post-processed with a high-pass filter (HPF) to remove smooth structures (atmospheric speckle halo), leaving the small scale high frequency components (e.g. planets) unaffected. These HPF images allow us to identify the limit imposed by the speckle background when the dynamic atmospheric wavefront has been reduced by the XAO system. Detectability estimates for profile evaluation are applied on HPF images, where contrasts evaluation are no longer suitable.

\subsection{Laboratory results}
\subsubsection{General trend}
\begin{figure*}
\includegraphics[width=5.cm]{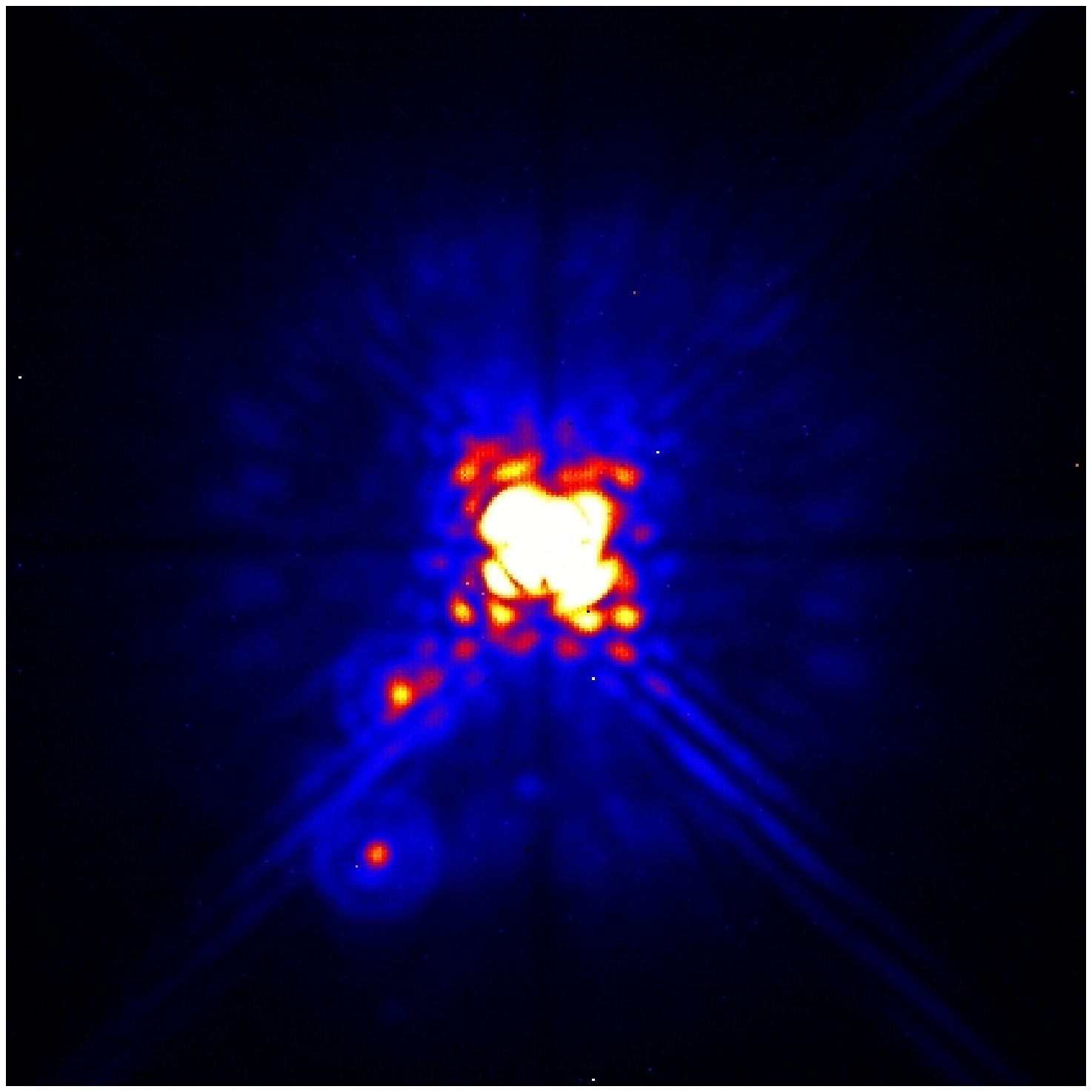}
\includegraphics[width=5.cm]{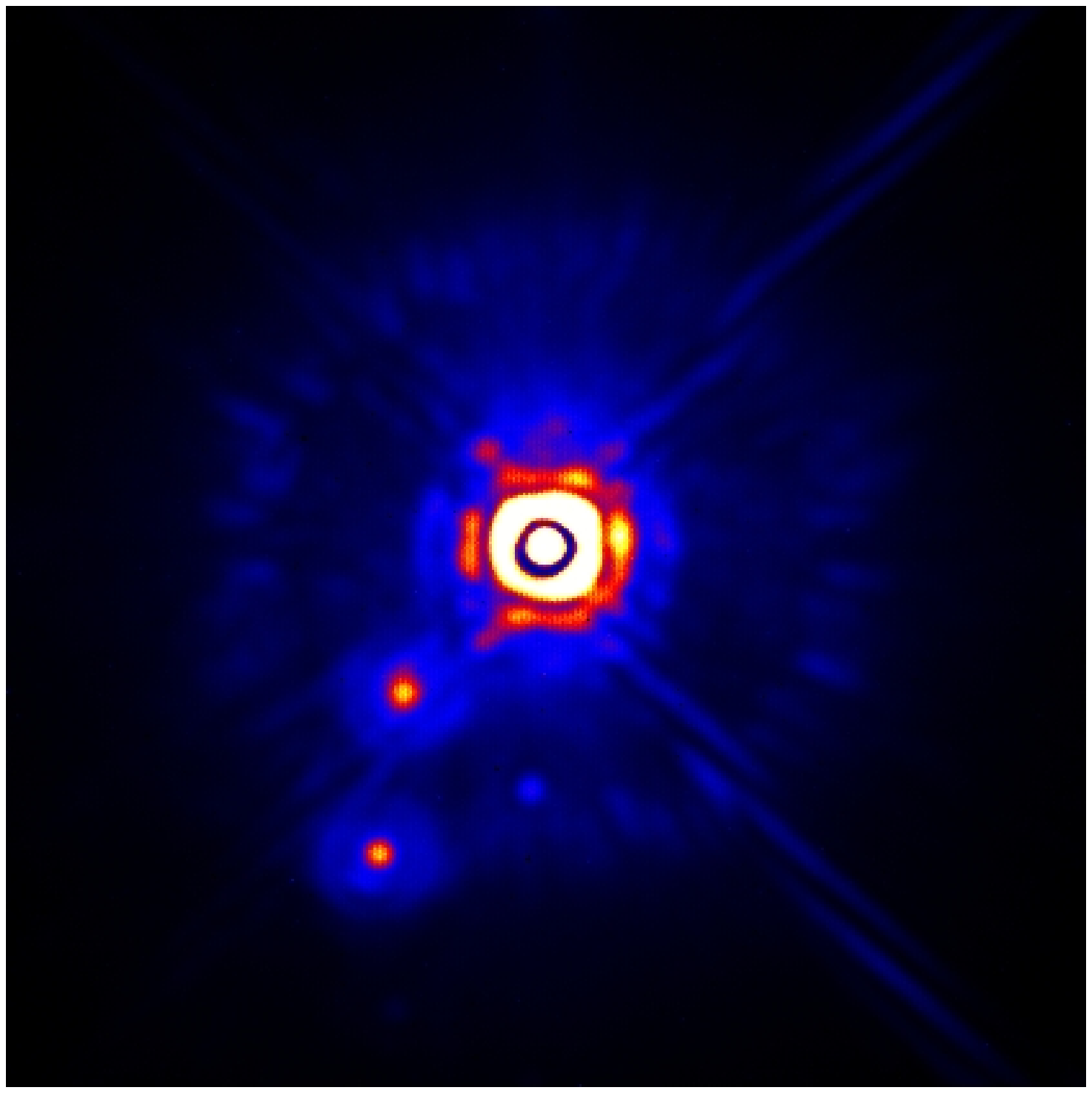}
\includegraphics[width=5.cm]{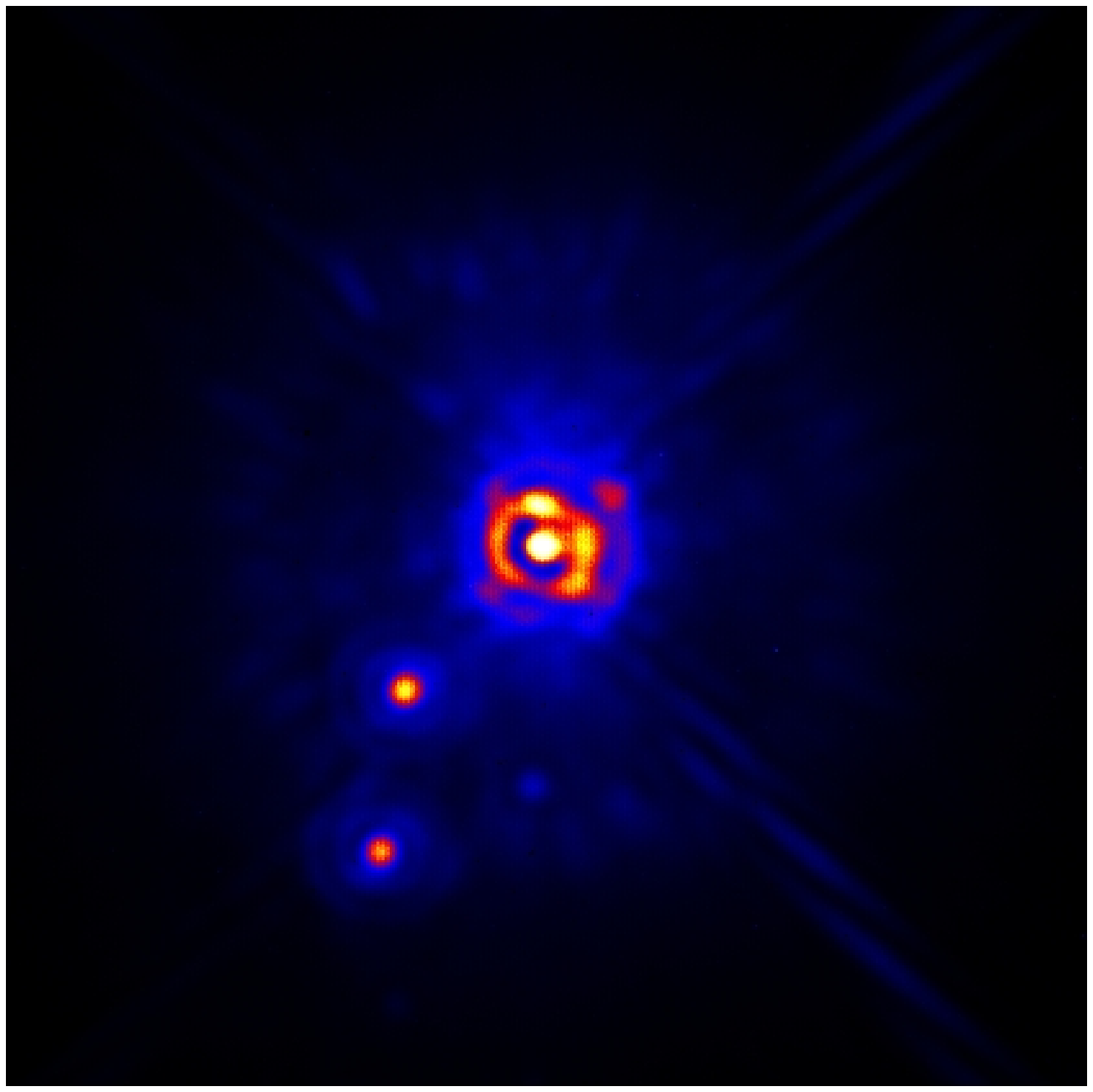}
\includegraphics[width=5.cm]{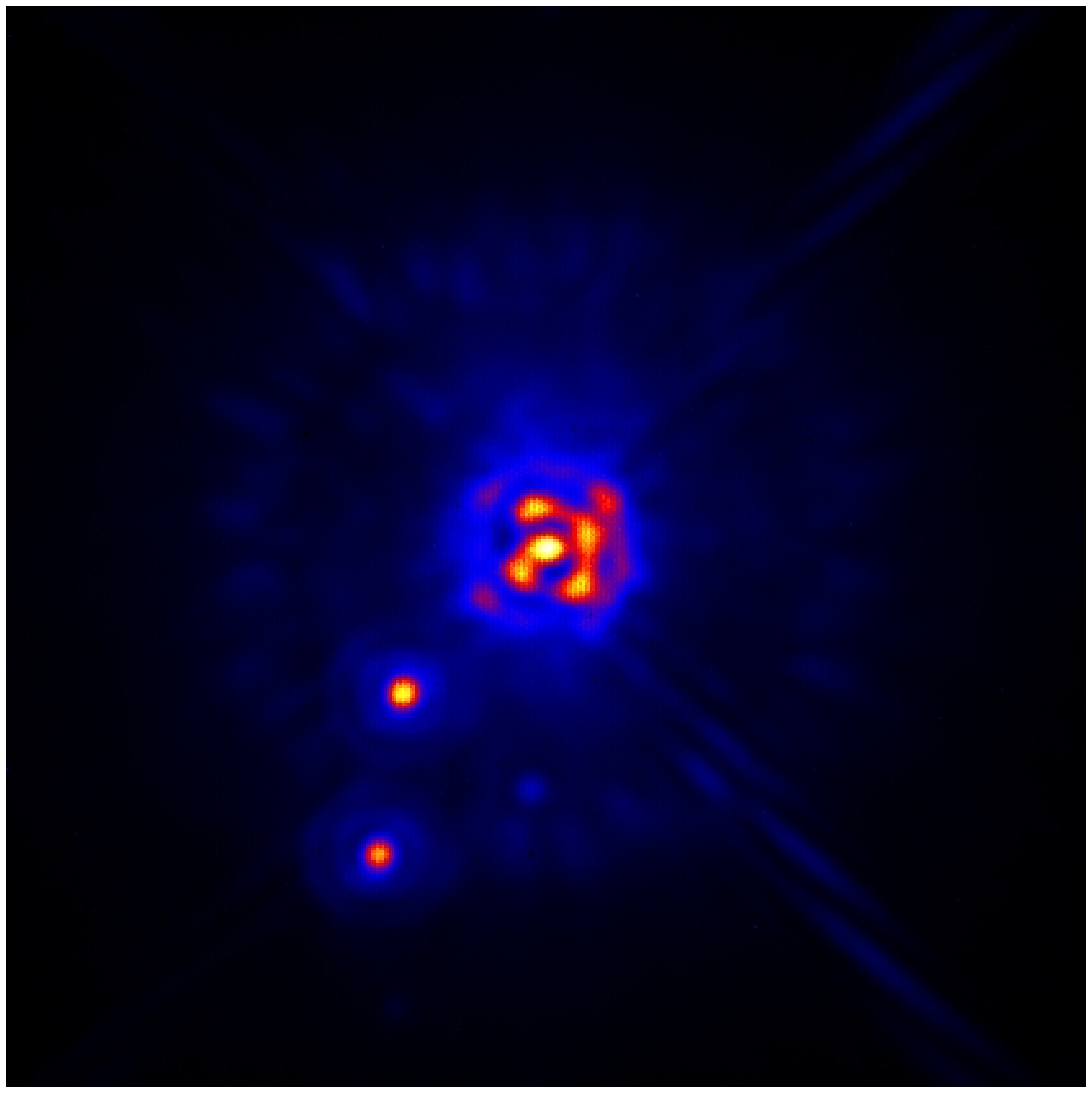}
\includegraphics[width=5.cm]{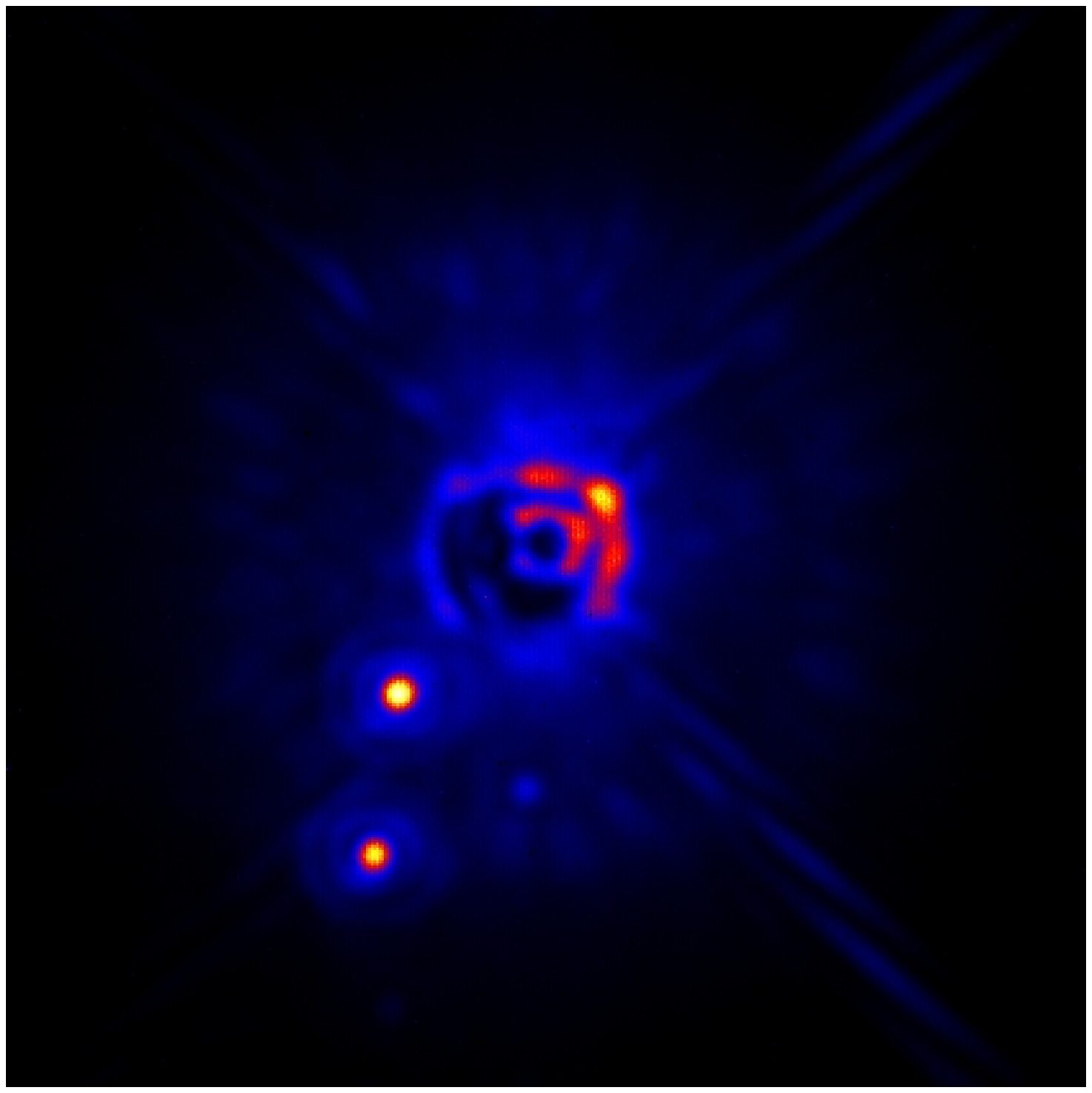}
\includegraphics[width=5.cm]{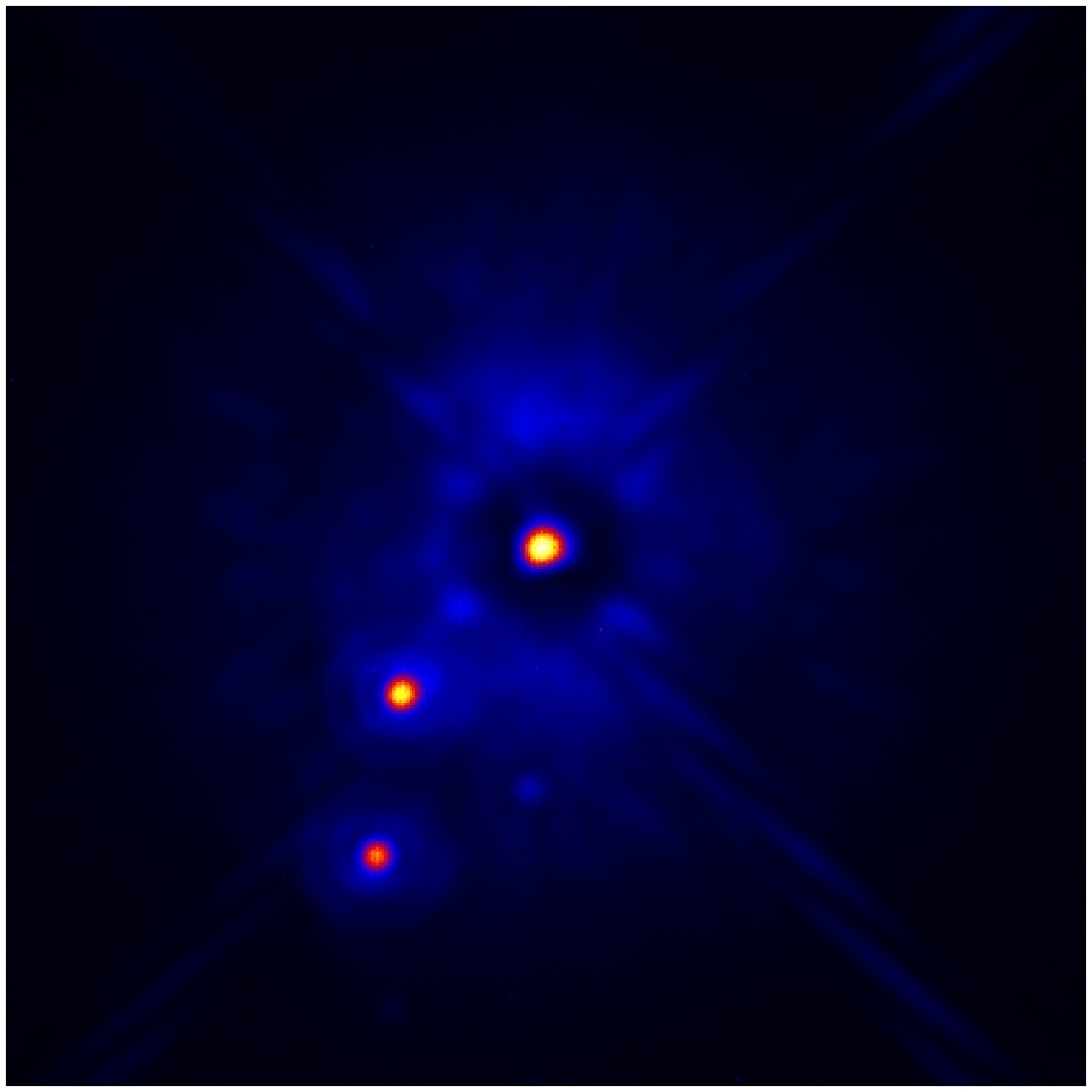}
\caption{XAO corrected coronagraphic images under 0.5$\arcsec$ seeing ($\Delta \lambda/\lambda = 1.4\%$). Top row (from the left to the right): FQPM, APLC, and LC(1) images. Bottom row (from the left to the right): LC(2), LC(3), and BLC images. The arbitrary color distribution and image dynamic aim at enhancing the contrast for the sake of clarity.}
\label{Images}
\end{figure*}

\begin{figure*}
\includegraphics[width=5.cm]{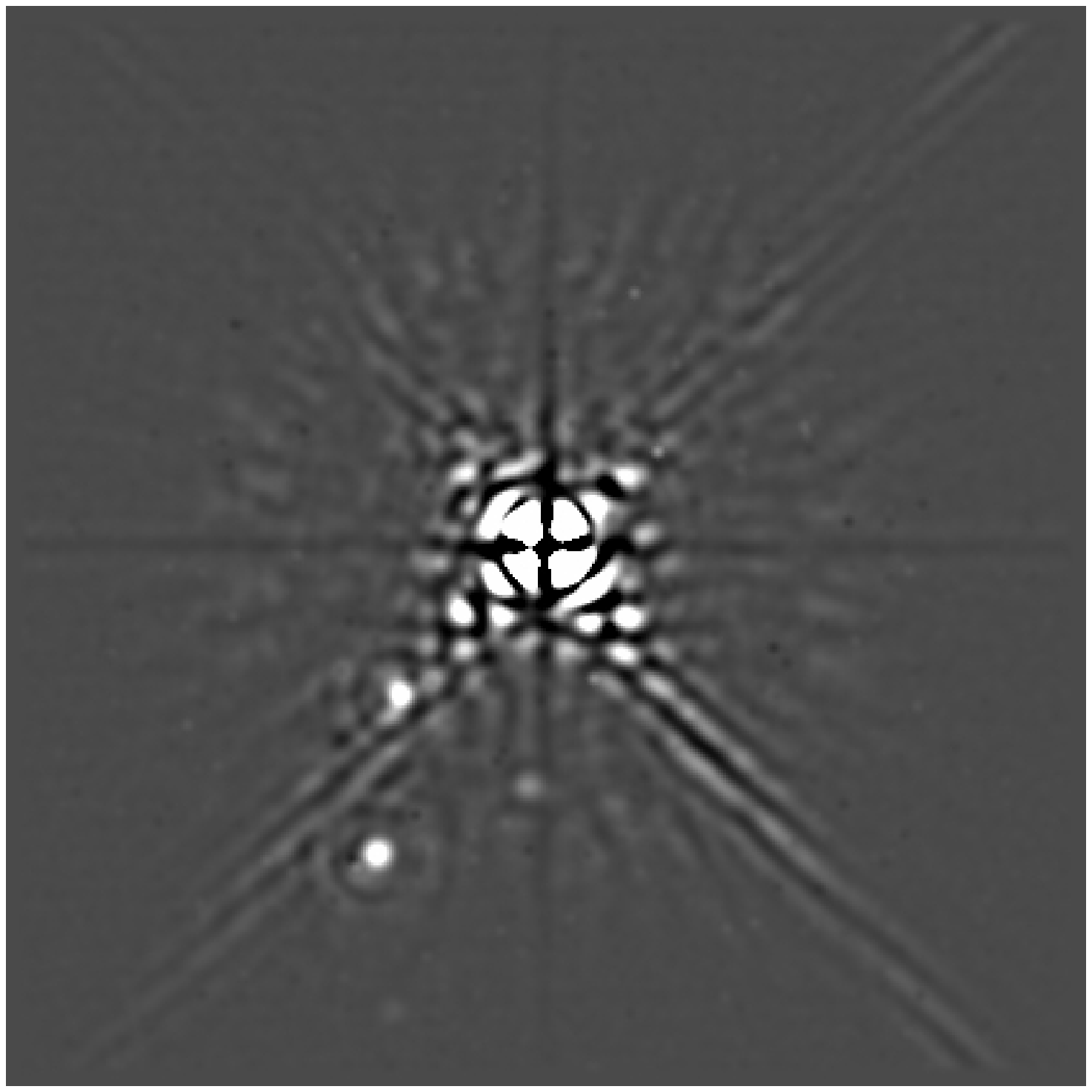}
\includegraphics[width=5.cm]{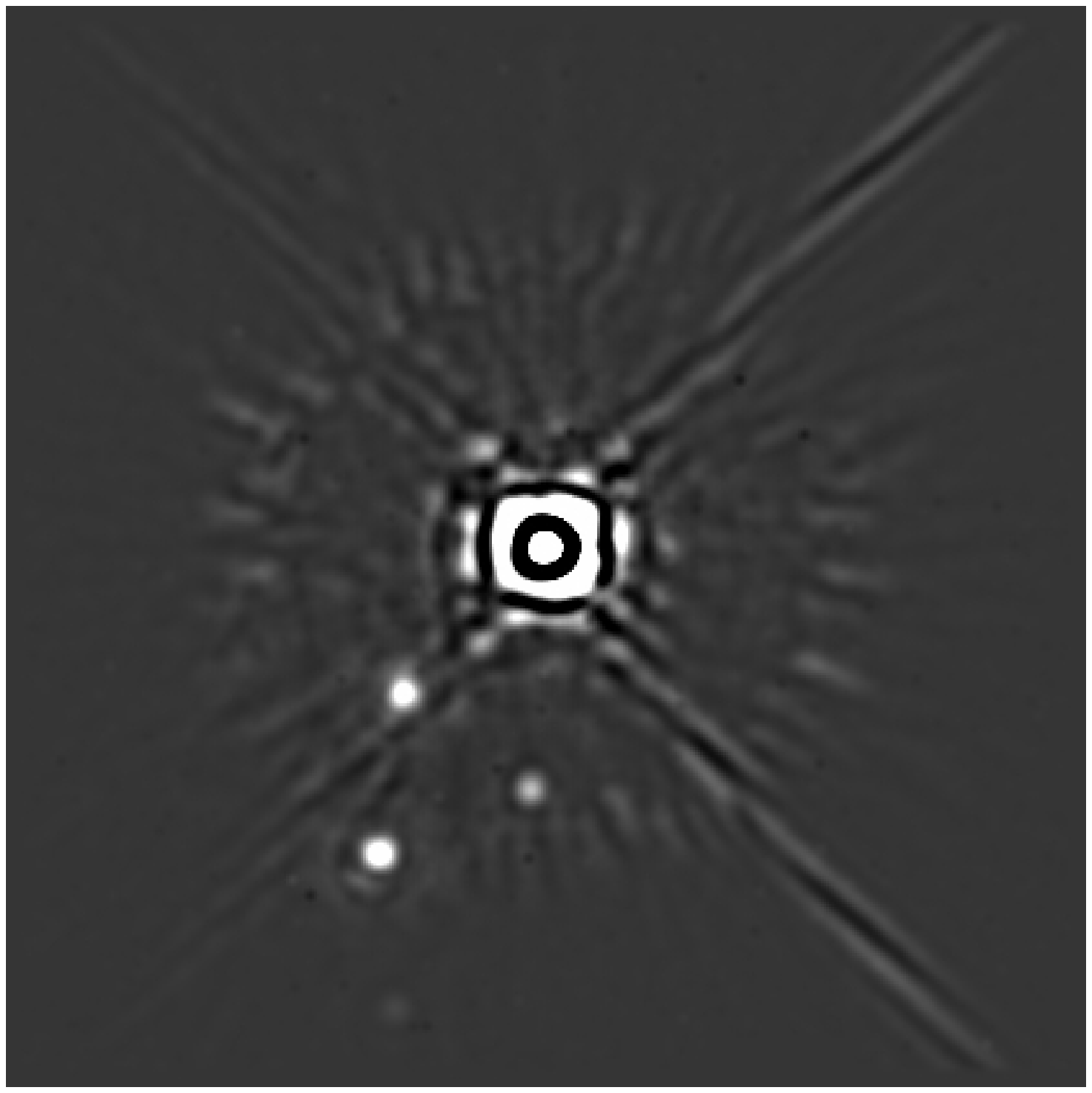}
\includegraphics[width=5.cm]{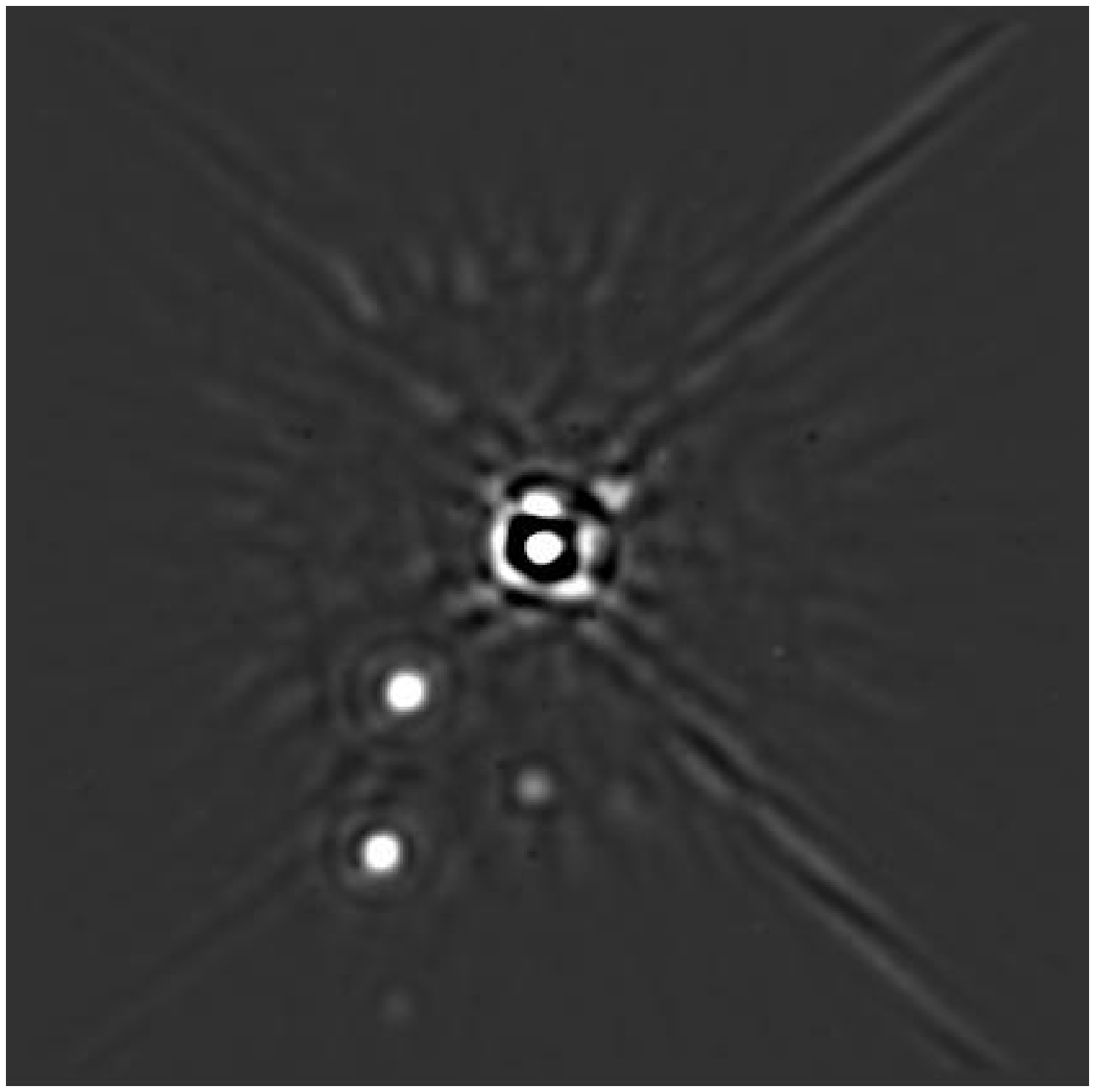}
\includegraphics[width=5.cm]{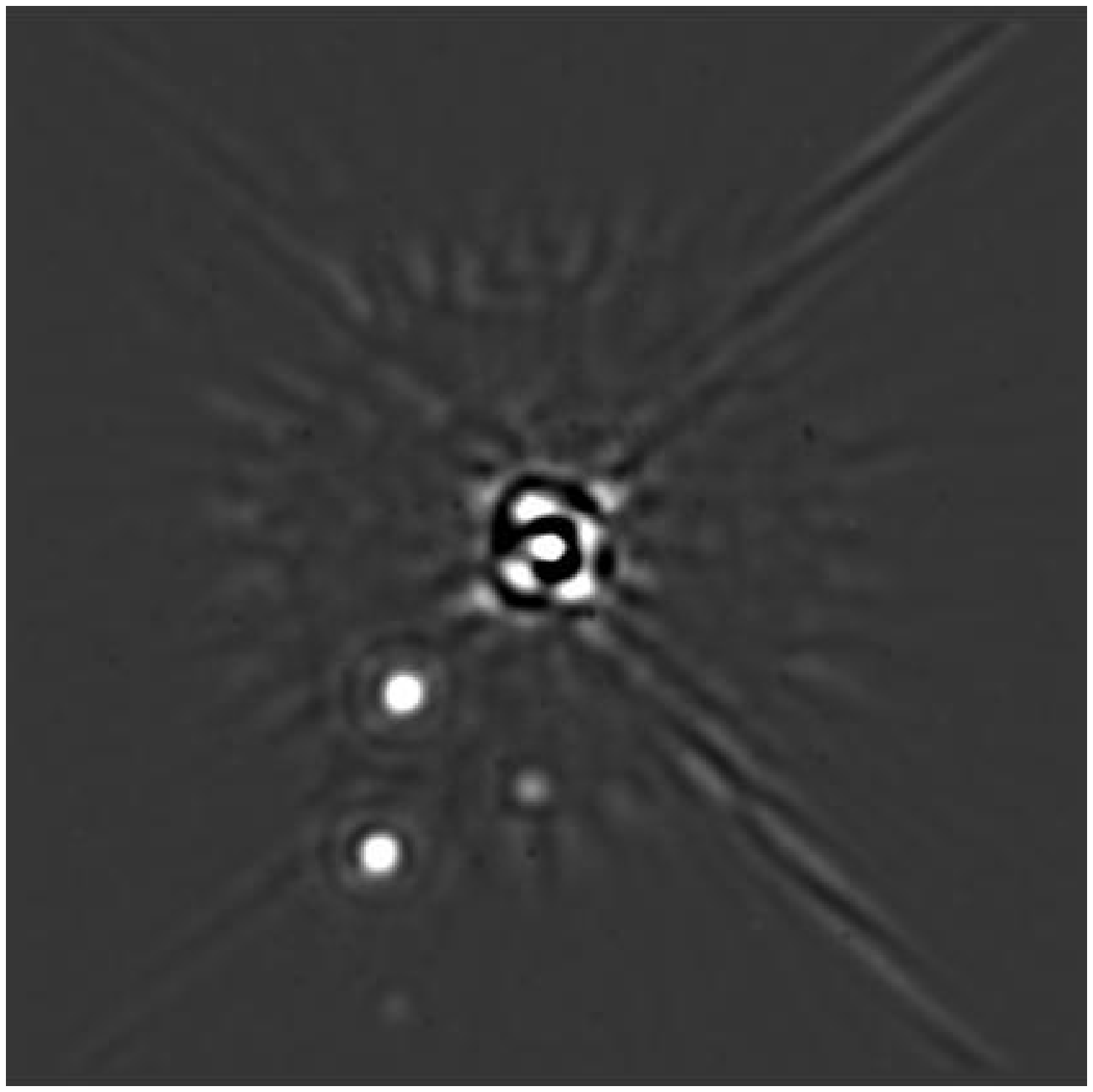}
\includegraphics[width=5.cm]{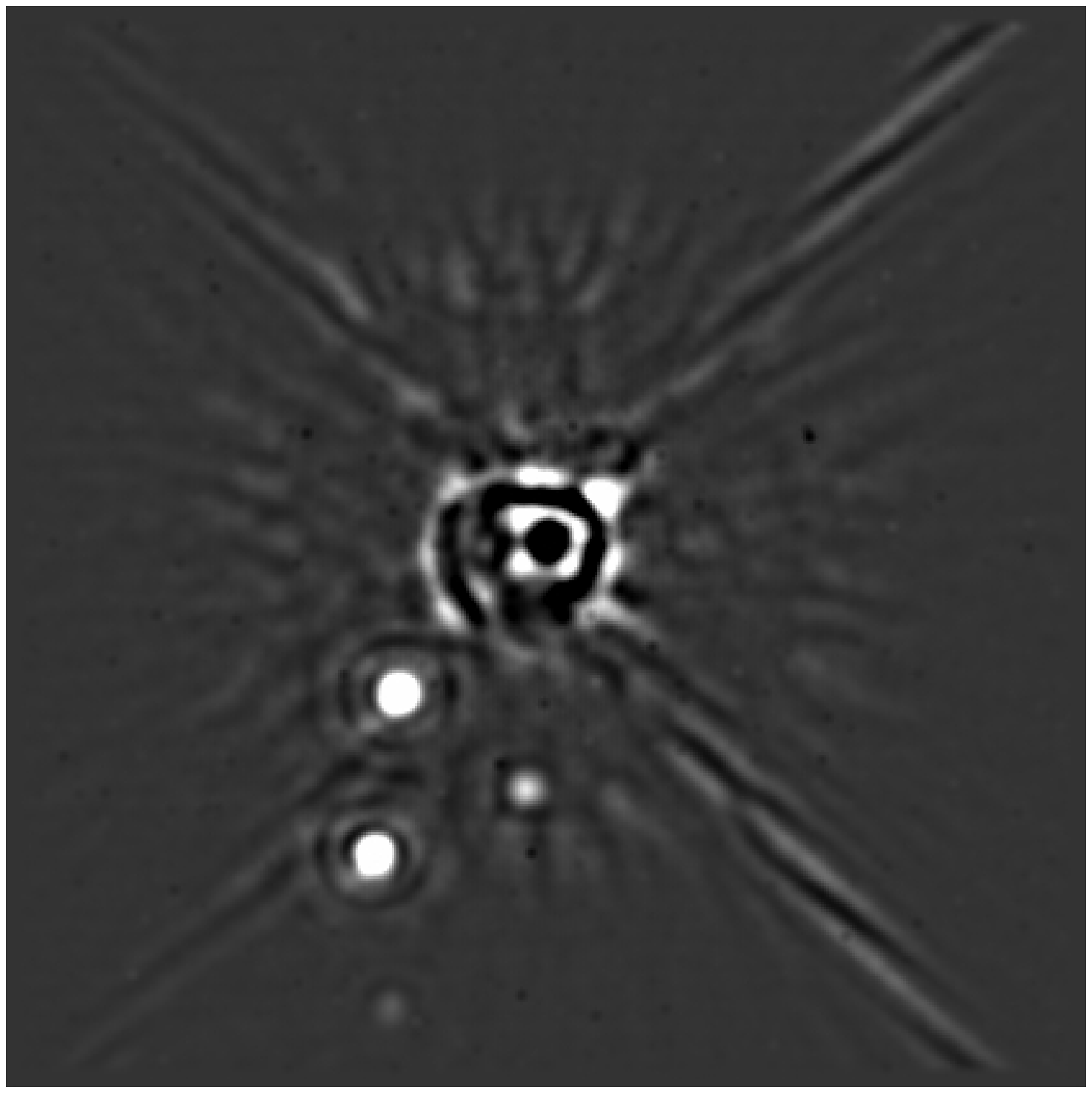}
\includegraphics[width=5.cm]{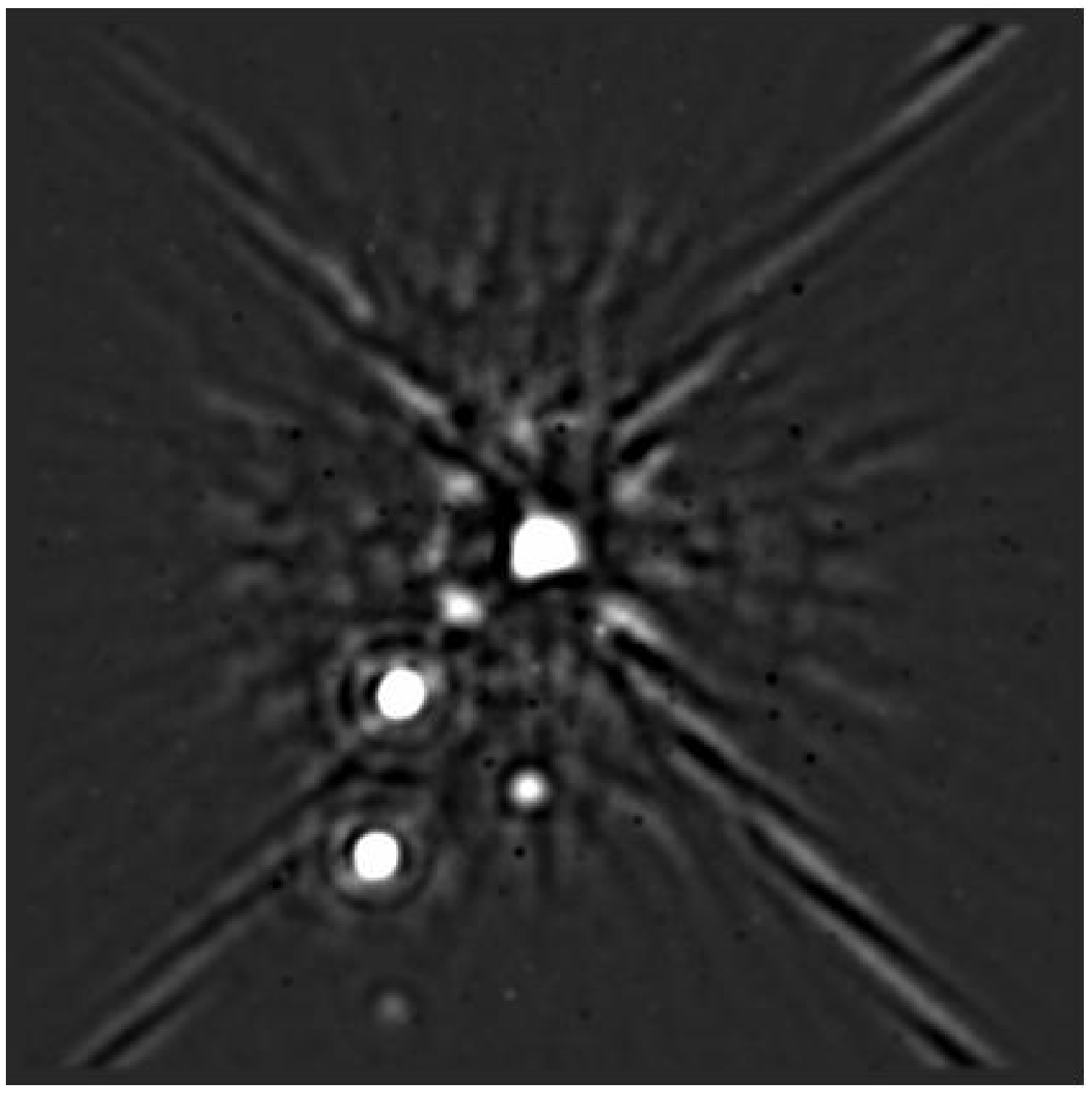}
\caption{High-pass filtered coronagraphic images (corresponding to Fig. \ref{Images}). Top row (from the left to the right): FQPM, APLC, and LC(1) images. Bottom row (from the left to the right): LC(2), LC(3), and BLC images. The arbitrary color distribution and image dynamic aim at enhancing the contrast for the sake of clarity.}
\label{Images2}
\end{figure*}

All raw images (Fig. \ref{Images}) present similar features: the coronagraphic images demonstrate starlight attenuation, and exhibit higher energy level at close angular separations (from the center of the image to basically the IWA) due to diffraction residuals and pinned speckles. A radial trend in speckle intensity is observable in the image: while speckles closer to the center are brighter, the speckle intensity decreases at larger angular separations until it increases to reach a local maximum at the AO cut-off frequency ($15\lambda/D$). The AO cut-off frequency is clearly observable in all images as its position in the field is identified by the slope of intensity in the speckle field at 0.6$\arcsec$ from the images center (as expected). \\
Diffraction resulting from the spider-vanes structures of the VLT-pupil is apparent in the images. The shape and extent of the central pattern of each coronagraphic image are influenced by the nature of the coronagraph (we remind the reader that IWA are different pending on the coronagraphic concept considered, see Table \ref{Resumtable}). For instance, the FQPM image 
exhibits the expected FQPM butterfly-like pattern with the four blind-zones introduced by the quadrant transitions on the mask. While the FQPM IWA is the sharpest one,  more energy is observable close to the star with this coronagraph due to the impact of the central obscuration. 
The HPF images (Fig. \ref{Images2}) correspond to coronagraphic images removed from smooth structures (i.e., atmospheric speckle halo), while small-scale high-frequency components remain alike.
\begin{figure*}
\includegraphics[width=8.5cm]{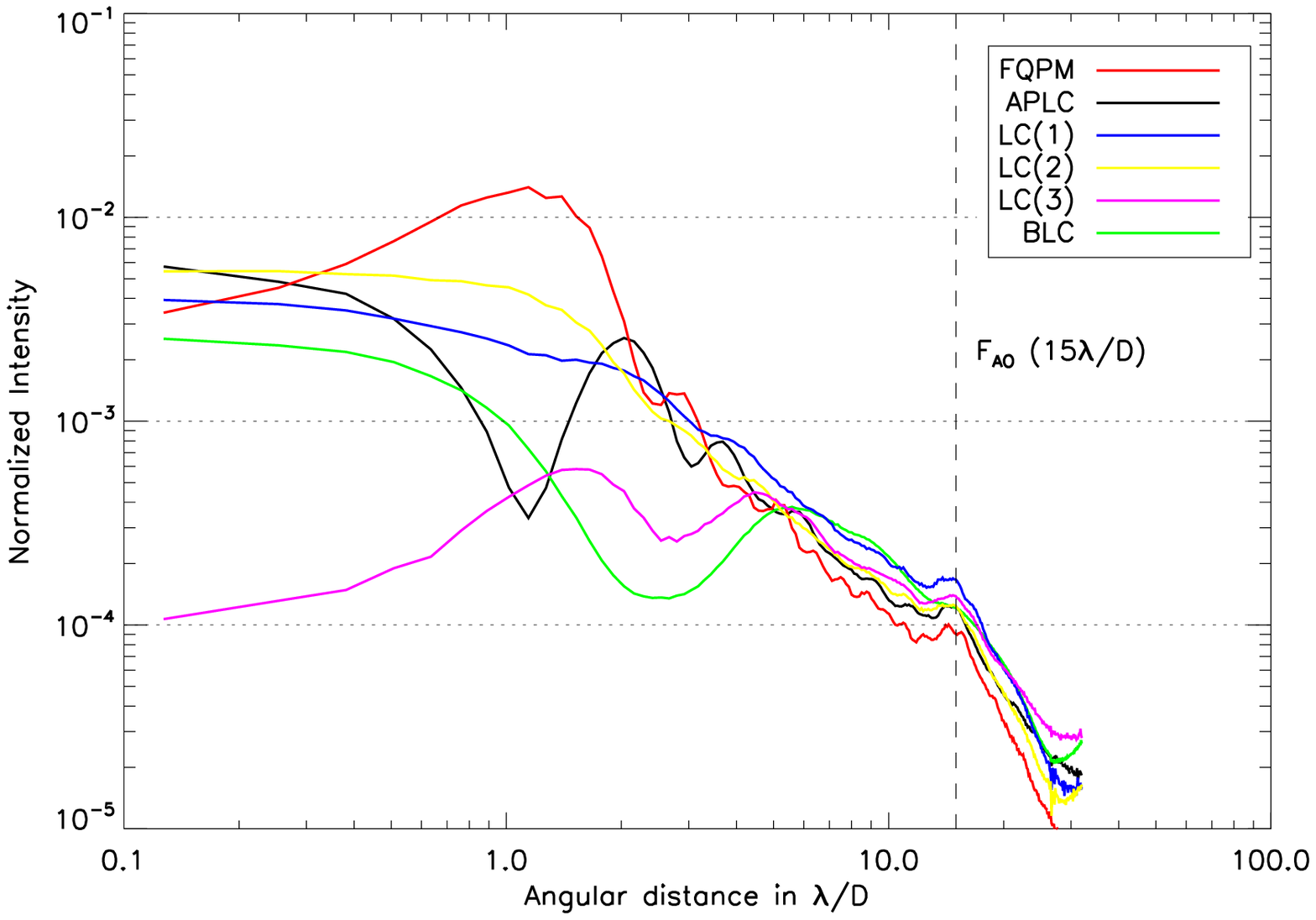}
\includegraphics[width=8.5cm]{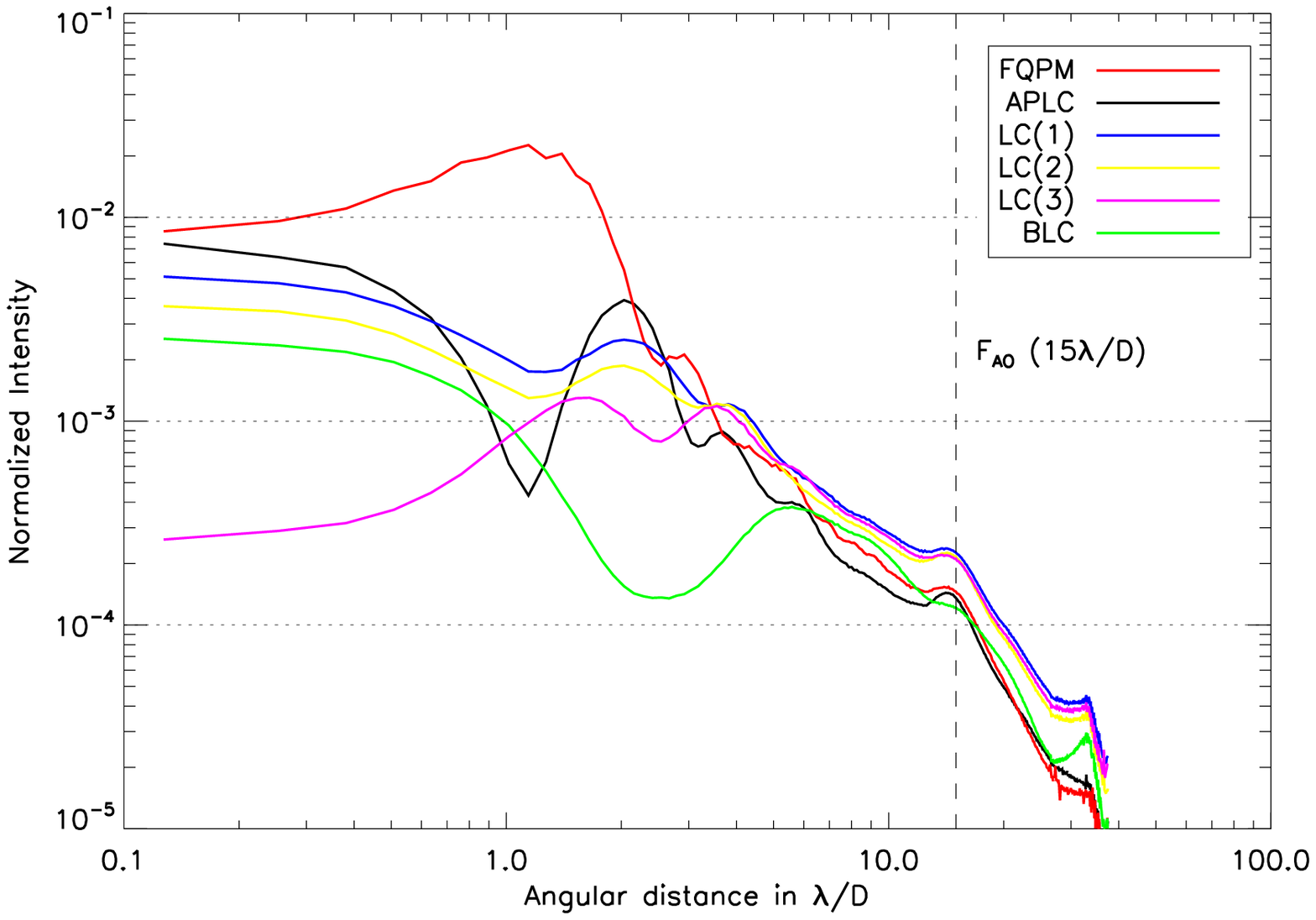}
\caption{Raw coronagraphic contrast profiles azimuthally averaged under 0.5$\arcsec$ dynamical seeing corrected by XAO (Strehl $\sim$90$\%$). Left: $\Delta \lambda/\lambda = 1.4\%$, Right: $\Delta \lambda/\lambda = 24\%$.}
\label{Results1}
\end{figure*} 
\begin{figure*}
\includegraphics[width=8.5cm]{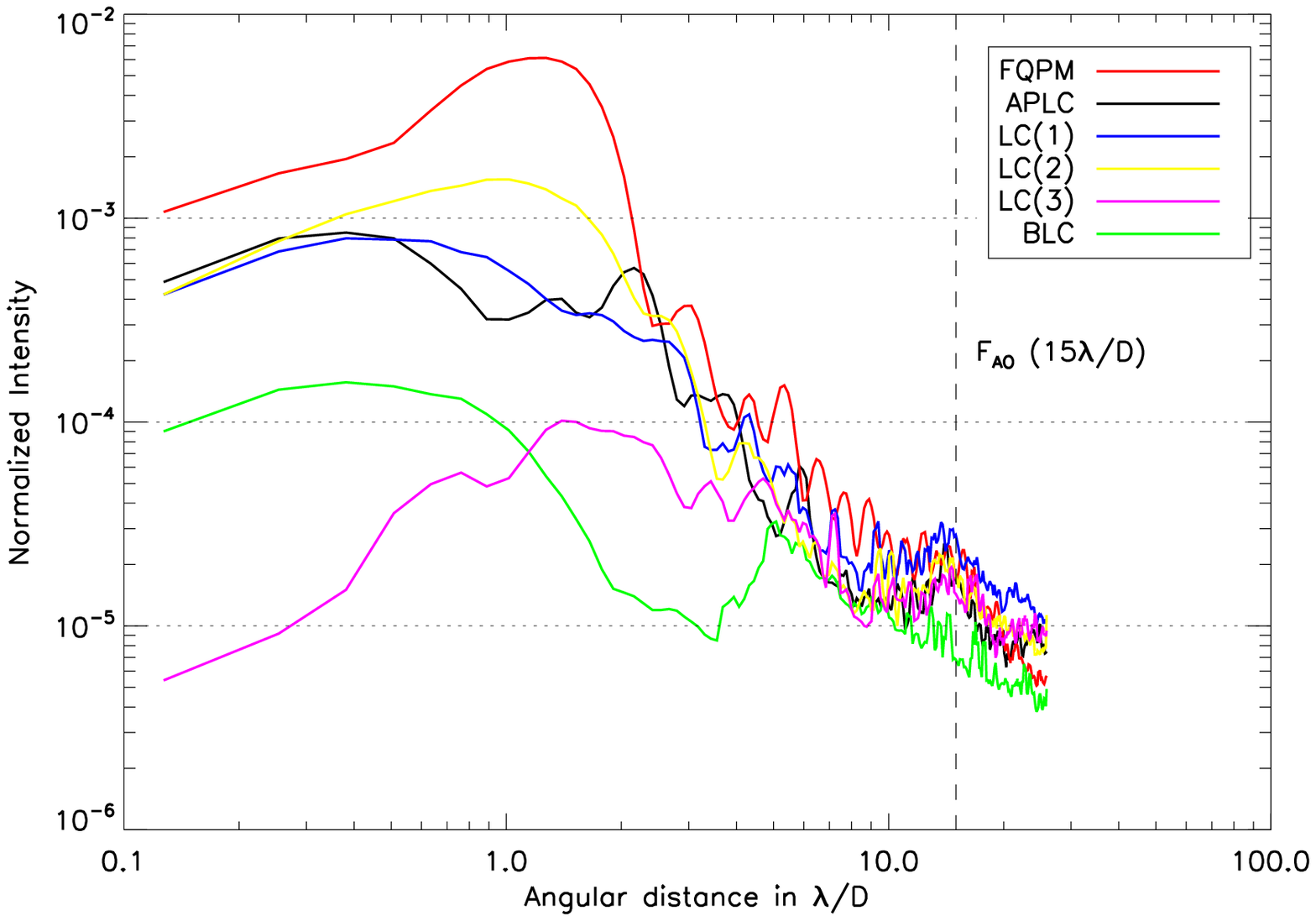}
\includegraphics[width=8.5cm]{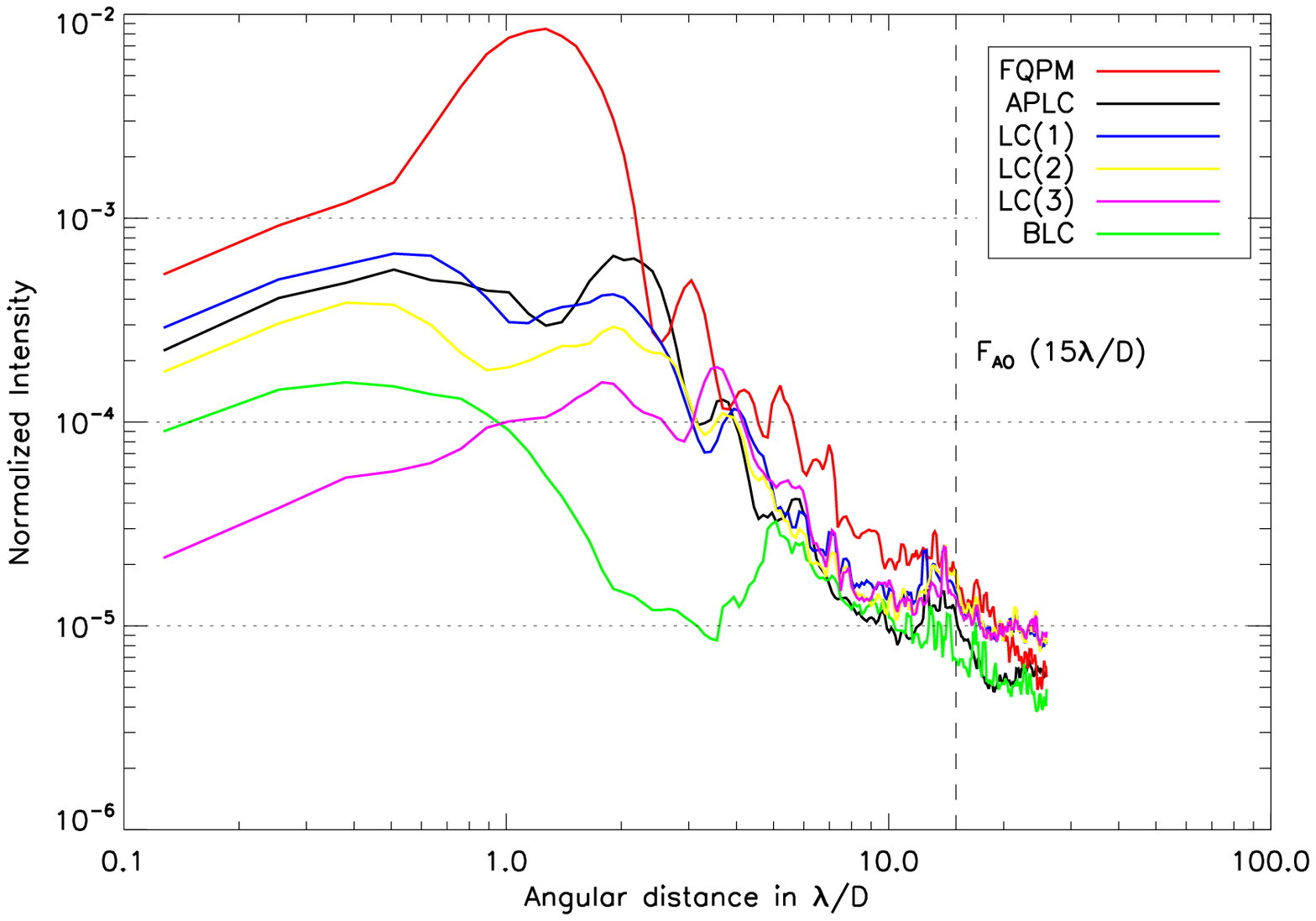}
\caption{High-pass filtered detectability profiles ($1\sigma$). Left: $\Delta \lambda/\lambda = 1.4\%$, Right: $\Delta \lambda/\lambda = 24\%$.}
\label{Results2}
\end{figure*} 
\begin{figure*}
\includegraphics[width=8.5cm]{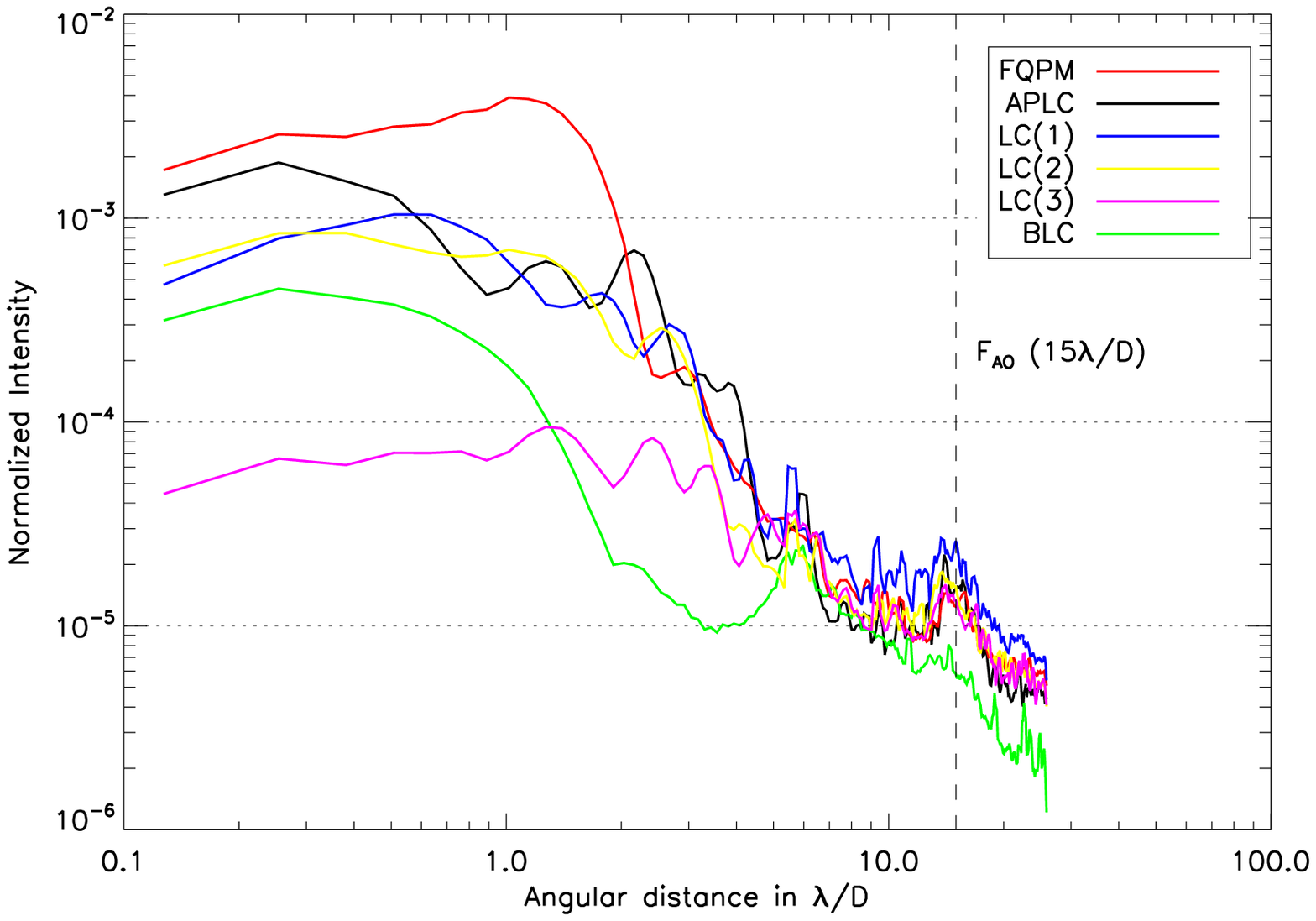}
\includegraphics[width=8.5cm]{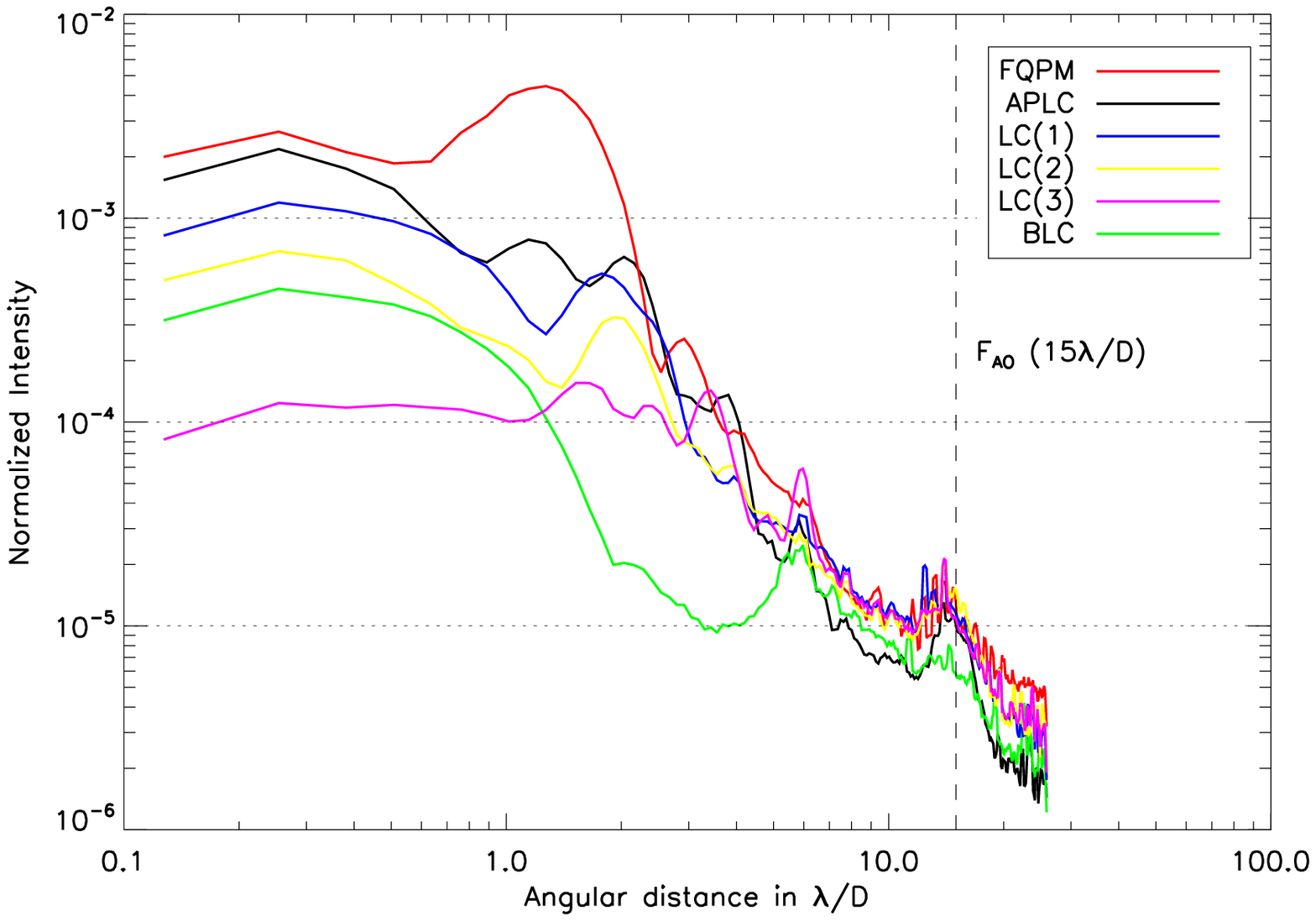}
\caption{High-pass filtered detectability profiles ($1\sigma$) evaluated in area free of spider diffraction pattern and ghosts (see Fig.  \ref{Area}).  Left: $\Delta \lambda/\lambda = 1.4\%$, Right: $\Delta \lambda/\lambda = 24\%$.}
\label{Results3}
\end{figure*} 

\subsubsection{Planet PSFs}
Four ghosts originating from reflections in the optical system before the coronagraph (indeed upstream of the IR-path of HOT, and potentially taking origin from the protecting window of the deformable mirror, or the beam splitter) are observable as bright PSFs in all images. None of them are observable in the images recorded on the coronagraphic testbench (Sect. \ref{PROTO}). Figure \ref{ghosts} (left) presents a close view of the APLC HPF image where the four ghosts (quoted A, B, C, and D) are observable. These ghosts are virtually similar in intensity in all coronagraphic images, although few differences are measurable due to the fact that coronagraph pupil-stops are not similar and modify the width of the PSFs. In addition the non-constant radial transmission of the BLC mask at large radii also introduce some differences. 
Ghost A, B, C, and D are localized in the field at 0.4$\arcsec$,  0.7$\arcsec$, 0.5$\arcsec$, and 0.9$\arcsec$, with peak intensity (1$\sigma$) of  $\sim$5 $10^{-4}$, $\sim$5 $10^{-4}$, $\sim$2 $10^{-4}$, and $\sim$3 $10^{-5}$ respectively.  

\subsubsection{Performance comparison}
All coronagraphs approximately deliver similar contrast levels from 5$\lambda/D$ to 15$\lambda/D$ (AO cut-off frequency, see Fig. \ref{Results1} and Table \ref{COROtable2}). Likewise, detectability obtained on HPF images is comparable (Fig. \ref{Results2}). Performance essentially differs from the center of the image to the IWA (different from a concept to another). Small IWA concepts (e.g. FQPM) deliver similar performance in the AO control domain as others, but the peak rejection rate is lower.  
While large IWA concepts (LC(3) or BLC) deliver very high peak rejection rate, contrast levels are not improved in the AO control domain. 
More standard IWA concepts (LC(1), LC(2), or APLC) stand in-between (improved peak rejection rate with respect to small IWA concepts, with similar contrast levels in the halo).   

From the total rejection obtained with the FQPM in narrow band and the budget error defined in Table \ref{FQPMtable} we can derive an estimation of the wavefront error after XAO (conservative assessment, i.e., all other error sources apart from the ones described in Table \ref{FQPMtable} are neglected). 
In the case of HOT the size on the source is 8$\mu$m (0.1$\lambda/D$) instead of 6.6$\mu$m (coronagraphic testbench), then the total rejection imposed by the stellar diameter changes from 1380 to 930.
We obtain an estimate of the wavefront error after XAO to be of the order of $\lambda/18$ rms at 1.64$\mu$m, i.e., $\sim$90 nm rms. This value is consistent with the 90$\%$ Strehl estimation at the same wavelength, i.e., $\sim$$\lambda/20$ rms, and reads realistically considering the 50 nm rms static aberrations measurement of the common path (left uncorrected by the 60-bimorph DM), and the 24 nm rms static aberrations estimation of the non-common path (IR-path, high-frequency wavefront error prior to the pupil-stop). A rough estimation of the AO residual wavefront error would therefore be at the level of $\sim$16 nm rms.\\
When the atmospheric residual halo is removed by post-processing (HPF images, Fig. \ref{Results2}), the detectability levels obtained either almost achieved the intrinsic limitation of the coronagraphs (a factor of 2 to 6 for the FQPM, or the APLC), or roughly attained the coronagraph limitation (LCs). The BLC case is particular as more than an order of magnitude difference remains between HPF contrasts and the actual limits imposed by the BLC prototype. These results indicate that an improvement in the wavefront error control would be advantageous for the BLC, while only a slight improvement can be obtained with the FQPM and APLC, and none with the LCs.

\begin{figure}
\centering
\includegraphics[width=4.0cm]{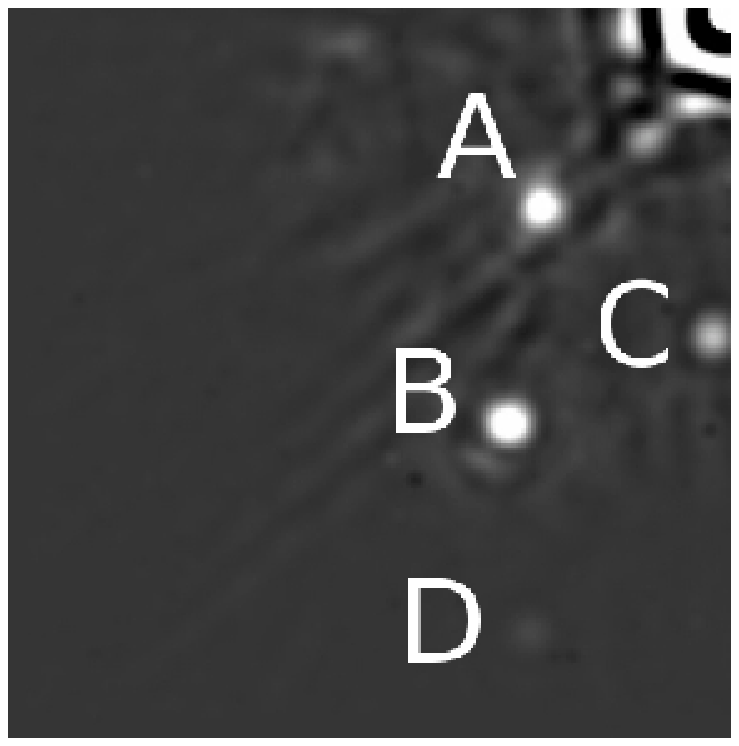}
\includegraphics[width=4.05cm]{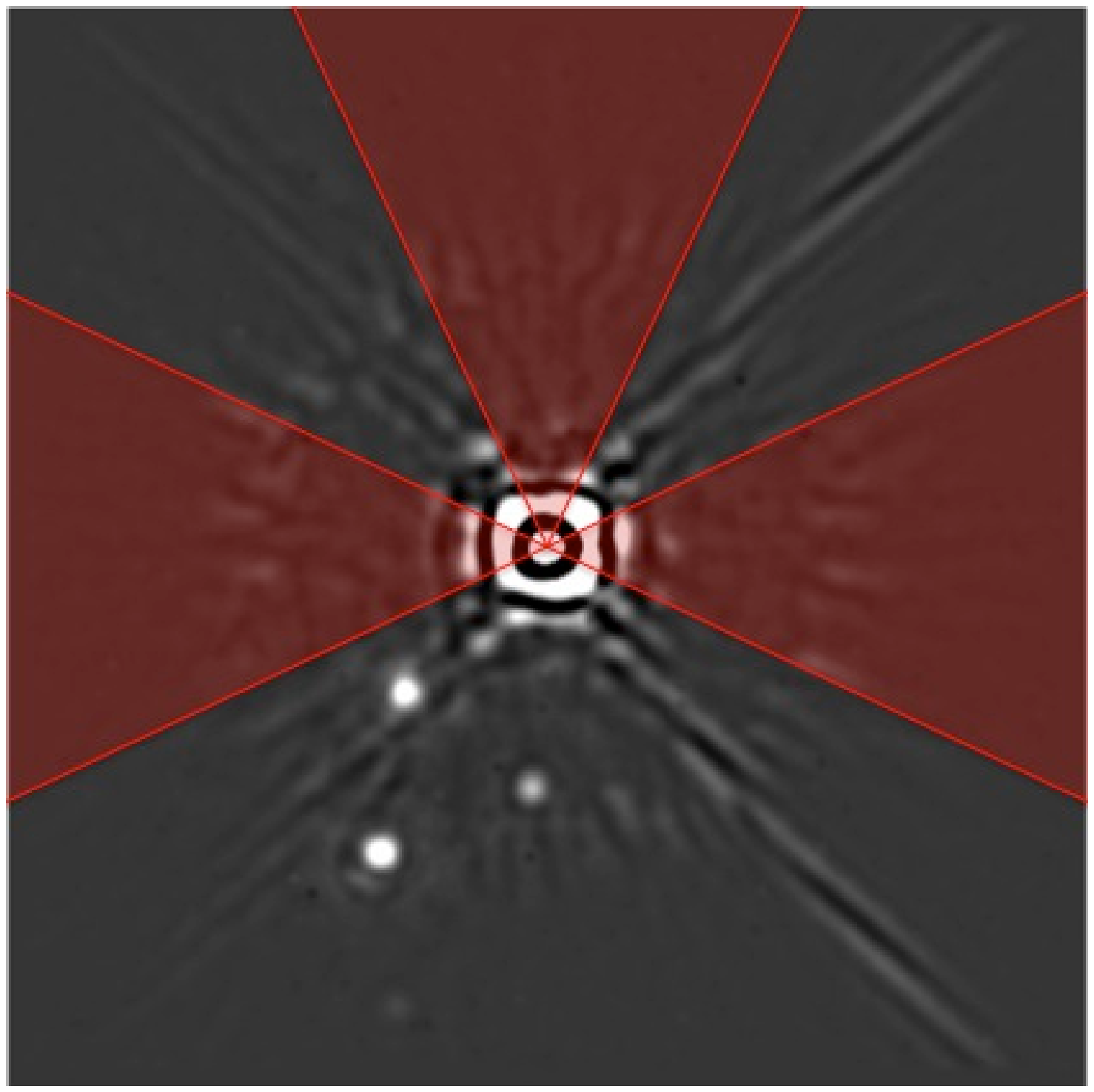}
\caption{Left: bottom left quarter of the APLC HPF image presented in Fig. \ref{Images2}, where the four ghosts in the image are identified (A, B, C, and D). Right: areas free of the spider quasi-static pattern and ghosts considered in the HPF images for the evaluation of the detectability presented in Fig. \ref{Results3}. }
\label{ghosts}
\end{figure}

As the azimuthal standard deviation (detectability) is a conservative estimate and the HPF images do not exhibit azimuthally uniform speckle noise distribution (i.e., spider vane residual structures are observable) more favorable detectabilities than the one presented in Table \ref{COROtable2} are locally evident. Therefore, by analyzing the HPF images in favorable areas that exclude remnant spider vane patterns (Fig. \ref{ghosts}, right) we can derive uppermost detectability levels reachable in each HPF image. Results are gathered in Fig. \ref{Results3} and demonstrate an improvement of the detectability that is a function of the angular separation and the coronagraph.
Basically, the improvement evolves from a factor 2 to 4.5 (from 3 to 12$\lambda/D$), and can nearly differ by a factor of 1.5 from a concept to another.  Finally, it brings the agreement of performance between all coronagraphs even more closer than previously presented.

Additionally, because we obtained lower performance than the one presented in Table \ref{COROtable2} with the BLC when combined with its pupil-stop (Fig. \ref{STOPS}, right), the results presented here correspond to the BLC used with the LC pupil-stop (less aggressive, Fig. \ref{STOPS}, second pattern from the left). In such situation, a contrast improvement by a factor varying from 1.3 to 2.1 was obtained in the halo, while the gain in throughput is about 18$\%$. This indicates that in real situation the use of an aggressive pupil stop with a BLC can be avoided even with a sophisticated entrance pupil (at least in a Strehl ratio $\leq$90$\%$ regime). This demonstrates the importance of a well-balanced error budget when designing complex systems. \\
The chromaticity impact is found negligible in the AO control domain, influencing contrast levels at minimum by a factor below 2 to a maximum of 5. Nearly all the impact is found in the area confined between the central core of the PSF and the IWA (e.g., the peak rejection rate, see Table \ref{COROtable2}).    
\begin{center}
\begin{table}
\centering
\begin{tabular}{l|l|l|l|l|l|l|}
\hline \hline 
 & BW ($\%$) & $\tau$ & $\tau_0$ & $\mathscr{C}_{3\lambda/D}$ & $\mathscr{C}_{12\lambda/D}$ & $\mathscr{C}_{20\lambda/D}$ \\ 
\hline \hline 
 \multicolumn{7}{c}{Raw images} \\
 \hline \hline
FQPM & 1.4 & 29 & 62 & 1.5 $10^{-3}$ & 9.7 $10^{-5}$ & 3.8 $10^{-5}$\\ 
            &  24 &  16 & 41 & 2.2 $10^{-3}$ & 1.6 $10^{-4}$ & 5.8 $10^{-5}$\\ 
\hline
LC (1)   & 1.4 & 65 & 244 &  1.0 $10^{-3}$ & 1.6 $10^{-4}$ & 6.1 $10^{-5}$\\ 
            &  24 &  57 & 191 & 1.4 $10^{-3}$ & 2.3 $10^{-4}$ & 1.0 $10^{-4}$\\ 
 \hline
LC (2)   & 1.4 &   55 &  183 & 8.9 $10^{-4}$ & 1.2 $10^{-4}$ & 4.6 $10^{-5}$\\ 
            &  24 & 71&  268 & 1.2 $10^{-3}$ & 2.0 $10^{-4}$ & 8.7 $10^{-4}$\\ 
 \hline
LC (3)    & 1.4 & 184 & 1719 & -- & 1.3 $10^{-4}$ & 6.2 $10^{-5}$\\
            &  24 &  103 & 767 &--  & 2.1 $10^{-4}$ & 9.2 $10^{-5}$\\ 
 \hline
APLC & 1.4 & 69 & 174 & 6.6 $10^{-4}$ & 1.1 $10^{-4}$ & 4.6 $10^{-5}$\\ 
            &  24 & 46 & 118 & 1.4 $10^{-3}$ & 1.3 $10^{-4}$&  5.4 $10^{-5}$\\ 
 \hline
BLC    & 1.4 & 136 & 373 & -- & 1.4 $10^{-4}$ & 6.5 $10^{-5}$\\ 
            &  24 & 140 & 373 & -- & 1.6 $10^{-4}$ & 6.5 $10^{-5}$\\ 
\hline \hline
 \multicolumn{7}{c}{High-Pass Filtered images } \\
 \hline \hline
FQPM & 1.4 &  -- & -- & 4.6 $10^{-4}$ & 2.3 $10^{-5}$ & 1.1 $10^{-5}$\\ 
            &  24 &  -- &  -- & 5.6 $10^{-4}$ & 2.8 $10^{-5}$ & 1.3 $10^{-5}$\\ 
\hline
LC (1)   & 1.4 & -- &  -- & 2.1 $10^{-4}$ & 2.3 $10^{-5}$ & 1.1 $10^{-5}$\\ 
            &  24 &  -- & -- & 1.3 $10^{-4}$ & 1.4 $10^{-5}$ &  9.1 $10^{-6}$\\ 
 \hline
LC (2)   & 1.4 & -- & --  & 2.2 $10^{-4}$ & 1.1 $10^{-5}$ & 1.1 $10^{-5}$ \\
            &  24 & -- & -- & 1.5 $10^{-4}$ & 1.3 $10^{-5}$ & 8.5 $10^{-6}$\\ 
 \hline
LC (3)    & 1.4 &  -- & -- & -- & 1.1 $10^{-5}$ & 8.3 $10^{-6}$\\ 
            &  24 & -- & -- &  --  & 1.2 $10^{-5}$ & 8.3 $10^{-6}$\\
 \hline
APLC & 1.4 &  -- &  -- & 1.1 $10^{-4}$ & 1.2 $10^{-5}$ & 8.5 $10^{-6}$\\ 
            &  24 &  -- &  -- &  1.5 $10^{-4}$&  1.1 $10^{-5}$ & 5.1 $10^{-6}$\\ 
 \hline
BLC    & 1.4 &  -- & --&  -- & 9.0 $10^{-6}$ & 4.2 $10^{-6}$\\ 
            &  24 &  -- & --&  -- & 8.9 $10^{-6}$ & 4.8 $10^{-6}$\\ 
\hline 
\end{tabular}
\caption{Performance of each coronagraph concepts obtained when tested on HOT w/ turbulence and XAO.}
\label{COROtable2}
\end{table}
\end{center}
\section{Conclusion}
This paper studies the efficiency of several coronagraphic concepts under realistic conditions and provides insights on their use in the context of forthcoming planet-finder instruments.
We selected and developed prototypes of various concepts (FQPM, LC, APLC, and BLC) and evaluated their performance under 0.5$\arcsec$ dynamical seeing corrected by an XAO system to a 90$\%$ Strehl ratio.\\
The delivered coronagraphic performances have been analyzed at two levels of contrast: \\
-- considering the residuals of the XAO system,\\
-- when this residual halo has been removed by post-processing.\\
In this context, similar conclusions are derived for both contrast levels regardless of the coronagraphic concept.
\begin{itemize}
\item At Strehl ratio $\sim$ 90$\%$ all coronagraphs deliver similar contrast levels from IWA to the AO cut-off frequency and are therefore suitable for such XAO performance. 
\item When the residual AO wavefront error is removed by post-processing, similar detectability is delivered by all coronagraphs.
\item As contrast levels are similar at larger angular separations than the IWA for all concepts, performance estimates for signal-to-noise predictions for real observations (exposure time calculation) with multi-coronagraph instruments (e.g., SPHERE) can in principle be achieved regardless of the coronagraphic concept. 
\item Increasing the IWA of a coronagraph (e.g., LCs, or BLC) does not provide better contrast, but increases the peak rejection. The peak rejection improvement varies from a factor $\sim$4 to $\sim$30 when the IWA increases from a factor 2 to 4 respectively (by comparing the FQPM to the LCs for instance). This aspect is important as wavefront errors downstream of the coronagraph produce quasi-static speckles that are proportional to the residual peak intensity. 
\item As small IWA is critical to probe the innermost region of astrophysical objects (basically only the FQPM offers the possibility to observe between 1 and 2 $\lambda/D$), efforts should be made to mitigate this level of residual energy in the center of the image by using, e.g., a FQPM combined with a small Lyot mask at its center, or by implementing mutli-FQPM designs \citep{baudoz07}.
\item At similar IWA the selection of a BLC over a LC should be restricted to situations where the XAO performance delivers better Strehl ratio than 90$\%$. 
This confirms the predictions from \citet{creep07}, where similar Strehl ratio regime was identified by means of simulations.
\end{itemize}

To take full advantage of a coronagraph the most demanding parameter is definitely the level of the wavefront control. The comparison of the coronagraphic contrast levels obtained on the two testbenches enables a differentiation for different levels of speckle noise, i.e., when the speckle limitation is set by the instrumental non-common path aberrations of HOT (coronagraphic testbench); and by either a combination of AO residual wavefront errors and instrumental common/non-common path aberrations (HOT, raw images), or common/non-common path aberrations only (HOT, HPF images).  A straightforward distinction between coronagraphs occurs only on the coronagraphic testbench where the uppermost concepts revealed are the APLC and BLC. 
This demonstrates the importance of a well-balanced error budget when designing complex systems. The conclusions of this experimental study support results from a previous analysis \citep{martinez08} comparing several coronagraphic concepts (including  the FQPM, LC, APLC, and the BLC) in the context of ELTs by means of numerical simulations. 

Speckle noise calibration/correction strategies are fundamental for direct detection of exoplanets with high-contrast instruments currently in use (Project 1640) or being commissioned (SPHERE, GPI, or HiCIAO), and were not addressed in this study. These instruments are foreseen to deliver better contrast levels (e.g., $10^{-7}$ to $10^{-8}$) than the one presented in this paper. 
To tackle the speckle noise limitation several solutions have been proposed based on spectral characteristics \citep{2000PASP..112...91M}, polarization states \citep{2003PASP..115.1363B}, image differential rotation \citep{2006ApJ...641..556M}, or coherence-based approaches \citep{2004ApJ...604L.117C, 2006dies.conf..553B, 2008A&A...488L...9G}. The dynamic range can be improved by orders of magnitude \citep[e.g.,][]{2006ApJ...641..556M, hinkley07, 2008ApJ...679.1574O, 2009ApJ...707L.123T, creep11}, while several authors have also proposed speckle subtraction methods through image post-processing \citep[e.g.,][]{2002ApJ...578..543S, 2006ApJ...641..556M, 2007ApJ...661.1208L}. In addition,  most of these instruments will take advantage of post-coronagraph wave front calibration systems which take aim at reconstructing the wavefront error at the critical location of the coronagraph \citep[e.g.,][]{sauvage07, 2008ApJ...688..701S, wallace10, hinkley11}.

\section*{Acknowledgments}
The activity outlined in this paper has been partially funded as part of the European Commission, Sixth Framework Programme (FP6), ELT Design Study, Contract No. 011863, Seventh Framework Programme (FP7), Capacities Specific Programme, Research Infrastructures, Contract number INFRA-2007-2.2.1.28, and 
Opticon joint research activity (JRA) 1 contract number RII3-CT-2004-001566.

\label{lastpage}

\end{document}